\documentclass[letterpaper,12pt]{article}
\usepackage[margin=1in]{geometry}
\usepackage{setspace}
\usepackage{parskip}
\usepackage{courier}
\usepackage{caption}
\usepackage{subcaption}

\usepackage{amsmath}
\usepackage{amssymb}
\usepackage{graphicx}
\usepackage{algorithm}
\usepackage{url}
\DeclareGraphicsExtensions{.pdf,.png,.jpg}

\def\IF{{\bf if}\ }

\def\ELSE{{\bf else}}
\def\WHILE{{\bf while}\ }

\def\blo{\noindent
\begin{tabular}{@{\quad}l@{\quad}}
\begin{minipage}{1in}
\begin{tabbing}
\qquad\=\qquad\=\qquad\=\qquad\=\qquad\=\qquad\=\qquad\=\kill}
\def\elo{\end{tabbing}\end{minipage}\\\end{tabular}}
\newenvironment{pseudocode}{\blo}{\elo}

\makeatletter
\newcommand*{\textoverline}[1]{$\overline{\hbox{#1}}\m@th$}
\makeatother
\newcommand{\cost}{\operatorname{cost}}
\newcommand{\size}{\operatorname{size}}

\newtheorem{theorem}{Theorem}%[section] uncomment to number theorems by section

\newtheorem{proposition}[theorem]{Proposition}

\begin{document}

\title{CAMP:  A Cost Adaptive Multi-Queue Eviction Policy for Key-Value Stores
\thanks{Sandy Irani and Jenny Lam are with the University of California, Irvine.  Their research is supported in part by the NSF grant CCF-0916181.  Shahram Ghandeharizadeh and Jason Yap are with the University of Southern California.}
\footnote{A shorter version of CAMP appeared in the Proceedings of the ACM/IFIP/USENIX 
Middleware Conference, Bordeaux, France, December 2014.  See \protect\url{https://github.com/scdblab/CAMP} for an implementation.}}

\author{
        {\em Shahram Ghandeharizadeh, Sandy Irani, Jenny Lam, Jason Yap } \\
%        \small Database Laboratory Technical Report 2014-07 \\
%        \small Computer Science Department, USC \\
%    \small Los Angeles, California 90089-0781 
}

%\date{}

\maketitle

\begin{abstract}
Cost Adaptive Multi-queue eviction Policy (CAMP) is an algorithm for a general purpose key-value store (KVS) that manages key-value pairs computed by applications with different access patterns, key-value sizes, and varying costs for each key-value pair. CAMP is an approximation of the Greedy Dual Size (GDS) algorithm that can be implemented as efficiently as LRU.  In particular, CAMP's eviction policies are as effective as those of GDS but require only a small fraction of the updates to an internal data structure in order to make those decisions. Similar to an implementation of LRU using queues, it  adapts to changing workload patterns based on the history of requests for different key-value pairs.  It is superior to LRU because it considers both the size and cost of key-value pairs to maximize the utility of the available memory across competing applications.  We compare CAMP with both LRU and an alternative that requires human intervention to partition memory into pools and assign grouping of key-value pairs to different pools.  The results demonstrate CAMP is as fast as LRU while outperforming both LRU and the pooled alternative.  We also present results from an implementation of CAMP using Twitter's version of memcached.   

\end{abstract}

%\subsection*{Abstract}

\section{Introduction}\label{sec:intro}
Applications with a high read-to-write ratio augment their persistent 
infrastructure with an in-memory key-value store (KVS) to enhance performance.  
An example is memcached in use by popular Internet destinations such as 
Facebook, Twitter, and Wikipedia.  Using a general purpose caching layer 
requires workloads to share infrastructure despite different access patterns, 
key-value sizes, and time required to compute a key-value 
pair~\cite{ugander11}.  An algorithm that considers only one factor may cause 
different application workloads to impact one another negatively, decreasing 
the overall effectiveness of the caching layer.  

As an example, consider two 
different applications of a social networking site:  one shows the profile of 
members while a second determines the displayed advertisements.  There may 
exist millions of key-value pairs corresponding to different 
member profiles, each computed using a simple database look-up that executed 
in a few milliseconds.  The second application may consist of thousands of 
key-value pairs computed using a machine-learning algorithm that processed 
Terabytes of data and required hours of execution.  This processing time is 
one definition of the \emph{cost} of a key-value pair. With a limited memory size and a high frequency of 
access for member profile key-value pairs, a simple algorithm 
that manages memory using recency of references
(LRU) may evict most of the key-value pairs of the second application,
increasing the incurred cost.

In general, reducing the incurred cost translates into a faster system that processes a larger number of requests per unit of time and may provide a better quality of service. The latter is due to availability of data (\emph{e.g.}, cache hit for a key-value computed using the machine learning algorithm) that enables the application to provide a user with more relevant content than content selected randomly. 
A possible approach is for a human expert to partition the available memory 
into disjoint pools with each pool managed using LRU. Next, the expert groups 
key-value pairs with similar costs together and assigns each group to a 
different pool~\cite{scalingmemcache}.  With our example, the expert would 
construct two pools.  One for the key-value pairs corresponding to members 
profiles and a second corresponding to advertisements.  The primary 
limitation\footnote{Partitioning is known to reduce the utilization of 
resources by resulting in formation of hot spots and bottlenecks.  One may 
address this limitation by over-provisioning resources.} of this approach is
that it requires a human familiar with the different classes of 
applications to identify the pools, construct grouping of key-value pairs, 
and assign each group to a pool.  Over time, the service provider may 
either introduce a new application or discontinue an existing one.  
This means the human expert must again become involved to identify the 
pool for the key-value pairs of the new application and possibly  
rebalance  memory across the pools once an application is discontinued. 

This paper introduces a novel caching method called Cost Adaptive Multi-queue 
eviction Policy (CAMP), that manages the available memory without 
partitioning it.  CAMP is an approximation of the Greedy Dual Size (GDS) 
algorithm~\cite{irani97} that processes cache hits and misses more 
efficiently using queues.  Hence, it is significantly faster than GDS and 
as fast as LRU.  It is novel and different from LRU in that it constructs 
multiple LRU queues dynamically based on the size and cost of key-value pairs. 
The number of constructed LRU queues depends on the distribution of costs and 
sizes of the key-value pairs.  CAMP manages these LRU queues without 
partitioning memory.  Thus, there is no need for human involvement to 
construct groups of key-value pairs, dictate assignment of groups to the 
pools, or configure and adjust the memory pool characteristics.   
CAMP is robust enough to prevent an aged expensive key-value pair 
from occupying memory indefinitely.  Such a key-value pair is evicted by 
CAMP as competing applications issue more requests.

CAMP is parameterized by a variable that controls its precision. At the highest precision,
CAMP's eviction decisions are essentially equivalent to those made by GDS. Our empirical results
show that CAMP does not suffer any degradation in the quality of its eviction decisions
at lower precisions. Moreover, it is able to make those decisions much more efficiently
than GDS. GDS requires an internal priority queue to determine a key-value pair to
evict from the cache. The time to maintain its data structures consistent in a thread-safe 
manner is expensive because it requires synchronization 
primitives~\cite{fekete13} with multiple threads performing caching decisions.
Moreover, CAMP performs a significantly fewer updates of its internal 
data structures than GDS, reducing the number of times it executes the
thread-safe software dramatically.

The rest of this paper is organized as follows.  Section~\ref{sec:camp} 
starts with a 
description of GDS to motivate CAMP and details its design decisions.  
Section~\ref{sec:eval} presents a simulation study of CAMP and compares it with LRU 
and the pooled approach that partitions resources, demonstrating its 
superiority.  Section~\ref{sec:impl} describes an implementation of CAMP using a 
variant of Twemcache and compares this implementation with the original 
that uses LRU.  Obtained results demonstrate that CAMP is as fast as LRU and provides
superior performance as it considers, in addition to recency of requests,
 the size and the cost of the
key-value pairs. Section~\ref{sec:related} describes related work. 
Section~\ref{sec:conc} provides brief words of conclusions and future research directions.

\section{CAMP}\label{sec:camp}
The algorithm Greedy Dual Size (GDS) was developed in the context of 
replacement policies for web proxy caches.  It captures many of the benefits of 
LRU and considers  the fact that data objects on the web have 
varying sizes and incur varying time delays to retrieve depending on their 
location and network traffic~\cite{irani97}. The same principles apply in the 
context of maintaining the identity of key-value pairs occupying the memory 
of a KVS, although the cost of an object in the KVS setting
may denote computation time
(or some other quantity) instead of retrieval time. 
Even though  the algorithm is applicable to a wide variety of settings, 
we adopt the terminology used for cache augmented database management systems~\cite{jason13}.
The GDS algorithm is based on an algorithm called Greedy Dual, developed by Neal 
Young~\cite{young91}, that handles objects of varying cost but uniform size. 
GDS assigns a value $H(p)$ to each key-value pair $p$ in the KVS. $H(p)$
is computed from a global parameter, $L$, as well as from
$\size(p)$, the size of $p$ and $\cost(p)$, the cost of $p$. The
value of $H(p)$
approximates the benefit of having that key-value in the KVS.
When there is 
insufficient memory to accommodate an incoming key-value pair, the algorithm 
continually evicts the key-value pair with the lowest value until there is 
room in the KVS to store the incoming key-value pair.

\begin{algorithm}
\begin{pseudocode}
Initialize $L \leftarrow 0$.\\
Process each request for key-value pairs in turn.\\
The current request is for key-value pair  $p$:\\
(1)\>\IF $p$ is already in memory (denoted by $M$),\\
(2)\>\> $L \leftarrow \min_{q \in M\setminus\{p\}} H(q)$.\\
(3)\>\ELSE,\\
(4)\>\> \WHILE there is not enough room in memory for $p$,\\
(5)\>\>\> Evict the $q$ with the smallest $H(q)$.\\
(6)\>\>\> $L \leftarrow \min_{q \in M} H(q)$. \\
(7)\>\> Bring $p$ into memory.\\
(8)\> $H(p) \leftarrow L+\cost(p)/\size(p)$.
\end{pseudocode}
\caption{GreedyDualSize.}
\label{alg:gds}
\end{algorithm}

The pseudocode for GDS is given in Algorithm~\ref{alg:gds}. 
The following proposition is useful in understanding how the algorithm works:

\begin{proposition}
~~
\begin{enumerate}
\item $L$ is non-decreasing in time.
\item If $p$ is in the KVS, then $L \leq H(p) \leq L + \cost(p)/\size(p)$.
\end{enumerate}
\end{proposition}

{\noindent{\bf Proof:}}
We prove the claim by induction on the number of requests.
At the beginning of the request sequence, there are no key-value pairs in the
KVS, so both claims are true. 
In lines 2 and 6, the value of $L$ is set to be the smallest $H$-value among
all the key-value pairs in the KVS. Since, by induction, for every key-value pair $p$ in the cache,
$L \le H(p)$, $L$ can only increase or stay the same after the change.
Lines 2 and lines 6 are the only time that $L$ changes, so the first
claim must hold. 

For claim 2, $L$ will not exceed $H(p)$ for any $p$ in the cache because its
new value (assigned in line 2 or 6)
is the smallest $H(p)$ among all the key-value pairs in the cache.
Any eviction performed in line $5$ only makes it easier to satisfy claim 2.
Finally, when a key-value pair is brought into the KVS, it's $H$ value is set to
$L + cost(p)/size(p)$ so the claim $2$ still holds by definition.
{\hspace*{\fill}\rule{6pt}{6pt}\bigskip}

\begin{figure}[ht]
        \begin{subfigure}{0.5\textwidth}
            \includegraphics{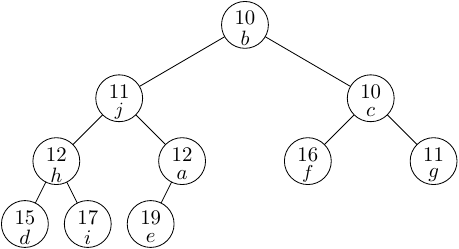}
            \caption{GDS's heap}
            \label{fig:heap-gds}
        \end{subfigure}\vspace{1em}
        \begin{subfigure}{0.49\textwidth}
            \hfill
            \includegraphics{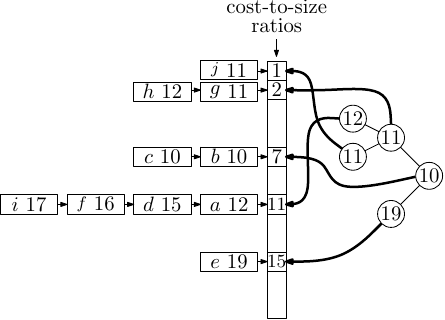}
            \caption{CAMP's priority queue}
            \label{fig:camp-gds}
        \end{subfigure}
        \caption{A heap used in a 
        straightforward manner (\ref{fig:heap-gds}) contains many more nodes 
        than the heap in CAMP's LRU-heap hybrid (\ref{fig:camp-gds}) when
        there are only a few distinct cost-to-size ratios.}
        \label{fig:gds-vs-camp}
\end{figure}

Consider a key-value pair $p$ in the KVS. As more key-value pairs are 
referenced (\emph{i.e.}, $p$ becomes requested less recently), the value of $L$ 
increases and $H(p)$ becomes smaller relative to the other key-value pairs in 
the KVS. If $p$ is requested while it is in the KVS, then $H(p)$ increases
to $L + \cost(p)/\size(p)$ which has the effect of delaying its eviction. All 
other things being equal, if a key-value pair has a \emph{cost-to-size ratio}, \emph{i.e.}, the quantity $\cost(p)/\size(p)$, that 
is $c$ times that of another key-value pair, it will reside in the KVS 
roughly $c$ times longer. GDS exhibits good performance under a variety of 
different load conditions because it considers both
varying costs and sizes without resorting to \emph{ad hoc} 
categorization of key-value pairs.

An implementation of GDS must maintain a data structure to identify and 
delete the key-value pair with the minimum priority efficiently.  
Typically, key-value pairs are maintained in a data structure that 
can retrieve and delete the key-value pair with the minimum priority. 
Normally this is accomplished by an implementation of a priority queue like a 
Fibonacci heap~\cite{Fredman87}. The worst-case performance of any priority 
queue is a $\log n$ cost per operation, where $n$ is the number of 
key-value pairs in the priority queue. For extremely large KVSs, such as 
those in use by Facebook and Twitter, overhead that scales even 
logarithmically as a function of the number of key-value pairs in the KVS 
results in a significant cost that could potentially be avoided. 

The starting point for this work is the observation that the $H$ value 
assigned to each key-value pair in the KVS is merely an approximation for the 
value of storing that key-value pair in the KVS in the absence of information 
about when that key-value pair will be requested next in the future relative 
to the other key-value pairs in the KVS. Insisting that the priority queue 
evict the key-value pair with the absolute minimum value is likely overkill. 
It seems reasonable that a similar KVS hit performance can be achieved if we 
only require that the data structure evict a key-value pair whose priority is 
only approximately smallest. An approximate priority queue could potentially 
be more efficient than one that is required to return the true minimum.

With GDS, the \emph{priority} or $H$ value (the two terms will be used interchangeably) of every key-value pair in the KVS is 
the sum of two values: the global non-decreasing variable $L$ and the 
cost-to-size ratio of the key-value pair. CAMP rounds the priority for every 
key-value pair by rounding the cost-to-size ratio before adding it to $L$. The 
rounding scheme results in a smaller set of possible cost-to-size values for 
key-value pairs stored in the KVS. CAMP takes advantage of the rounding by 
grouping the key-value pairs in its data structure according to the cost-to-size 
ratio instead of by priority value. 
The eviction decisions between CAMP and GDS differ slightly for two reasons. 
First, CAMP rounds the cost-to-size ratio in determining the $H$ 
value of a key-value pair. Second, in evicting the key-value 
pair with smallest priority, CAMP breaks ties according to
LRU, whereas GDS breaks ties arbitrarily. 

Figure~\ref{fig:gds-vs-camp} shows the
storage schemes for GDS and CAMP, with each circle denoting the $H$ value of a key-value pair. 
Figure~\ref{fig:heap-gds} shows a typical priority-queue-based implementation 
of GDS in which a set of key-value pairs are stored in a heap\footnote{A heap 
is a tree-based implementation of a priority queue which maintains the 
property that the priority of any node in the tree is at most 
the priority of its children.} based on their priority value. 
Figure~\ref{fig:camp-gds} shows CAMP's data structure in which 
key-value pairs are 
grouped into queues according to their cost-to-size ratio. Key-value pairs 
in a queue are ordered according to their priority which is their $H$ value. CAMP 
maintains a heap containing the priority of the data item at the head of every queue. 
Thus, to identify a candidate key-value pair to evict from the KVS, CAMP locates the key-value pair with the smallest priority among the heads of each of the queues.

CAMP's implementation is efficient based on the following key observation. If 
the key-value pairs within each queue are stored according to LRU, then the 
key-value pairs are automatically ordered according to their priority. For 
this reason, the queues maintained by CAMP are termed LRU queues. 
To understand why this observation holds, recall that all the items within an 
LRU queue have the same cost-to-size ratio. Furthermore, the $H$ value of a 
key-value pair is the value of $L$ at the time of its
last request plus its cost-to-size ratio. Since $L$ increases over time, a 
key-value pair that is requested 
earlier on will have a smaller $H$ value and appear towards the front of the 
queue, whereas a key-value pair that is referenced more recently will have a 
larger $H$ value and appear towards the end of the queue. In particular, the 
first key-value pair in each queue has the smallest priority.

\begin{figure}[h]
    \centering
    \includegraphics{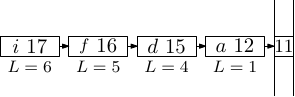}
    \caption{An LRU queue within CAMP.}\label{fig:lru}
\end{figure}

As an example, consider the LRU queue in Figure~\ref{fig:lru} which is one of 
the five queues shown in Figure~\ref{fig:camp-gds}. It points to the array 
value 11, containing all the key-value pairs that have cost-to-size ratio 11. 
Each node shows the key-value pair's $H$ value, and the shown $L$ value is 
the value of $L$ at the time of its last request. Since $L$ increases over 
time and key-value pairs are inserted at the tail of the queue, key-value 
pairs are ordered by the value of $L$ at the time of the request. Since all 
these key-value pairs have similar cost-to-size values, they are also ordered 
by their $H$ value. Specifically, $a$ is the least recently requested key-value 
pair in its queue.

With CAMP, the complexity of processing a KVS hit for a referenced key-value 
pair is the complexity to update the LRU queue ($O(1)$) plus the complexity 
to update the heap.
The worst case for the latter is logarithmic in the number of non-empty queues 
instead of the number of key-value pairs in the KVS since the nodes 
in the heap tree structure of CAMP identify LRU queues, see Figure~\ref{fig:camp-gds}.
With our implementation of CAMP, we chose to use an 8-ary implicit heap as suggested
by the recent study~\cite{LarkinST14} on priority queues. Here, 8-ary means with branching factor at most 8. Moreover, a heap is implicit if it uses the usual array implementation rather than with pointers.

\begin{figure}%[t]
    \begin{subfigure}{0.5\textwidth}
        \includegraphics{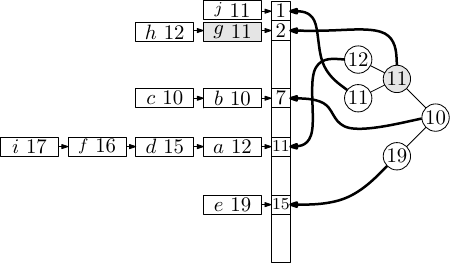}
        \caption{}
        \label{fig:camp-hit1}
    \end{subfigure}\vspace{1em}
    \begin{subfigure}{0.5\textwidth}
        \includegraphics{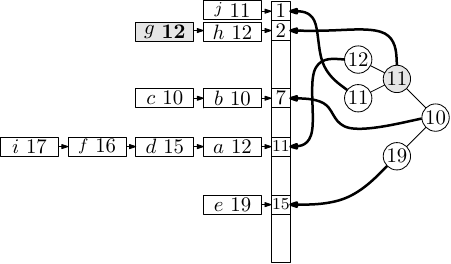}
        \caption{}
        \label{fig:camp-hit2}
    \end{subfigure}\vspace{1em}
    \begin{subfigure}{0.5\textwidth}
        \includegraphics{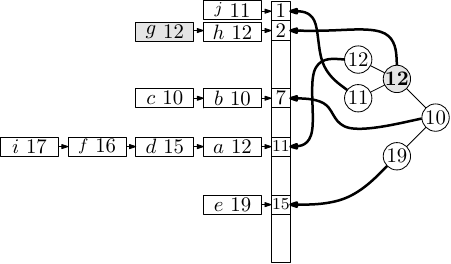}
        \caption{}
        \label{fig:camp-hit3}
    \end{subfigure}\vspace{1em}
    \begin{subfigure}{0.5\textwidth}
        \includegraphics{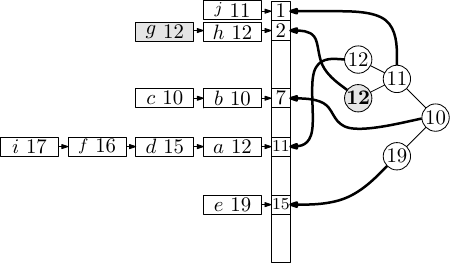}
        \caption{}
        \label{fig:camp-hit4}
    \end{subfigure}
    \caption{CAMP update on a KVS hit. If $g$ is requested (\ref{fig:camp-hit1}), it is moved to the back of its queue (\ref{fig:camp-hit2}), and the heap is updated accordingly (\ref{fig:camp-hit3} and \ref{fig:camp-hit4}).}\label{fig:camp-hit}
\end{figure}

To illustrate the processing of a KVS hit using the same running example of Figure~\ref{fig:gds-vs-camp},
consider a new reference for $g$.  
The KVS locates $g$ using a hash table (Figure~\ref{fig:camp-hit1}), moves it to the end of its LRU queue 
(see Figure~\ref{fig:camp-hit2}),
and updates its value to $10+2=12$ where 10 is the minimum priority ($L$ value)
and 2 is the cost-to-size ratio of $g$.
Now, the new head of the queue has priority 12.
CAMP updates the value of the heap node pointing to this queue (Figure~\ref{fig:camp-hit3}), causing the heap to be updated as shown in Figure~\ref{fig:camp-hit4}.

One way to improve performance is by limiting the number of LRU queues.  
We can do so by assigning key-value pairs with ``similar'' cost-to-size values to 
the same queue. Similarity in this context has a specific meaning, in that 
values that have different orders of magnitude should remain distinct. 
Therefore, a rounding scheme that simply truncates a fixed number of bits 
will not work. 
Instead, CAMP uses the integer rounding scheme described in~\cite{Matias96}. 
Given a number $x$, let $b$ be the order of its highest non-zero bit. To 
round $x$ to precision $p$, zero out the $b-p$ lower order bits or, in other 
words, preserve only the $p$ most significant bits starting with $b$. 
If $b \le p$, then $x$ is not rounded.
Table~\ref{table:rounding} illustrates the difference between these rounding 
schemes with examples. With regular rounding, too much information is kept 
for large values and too little information is kept for small values. Since 
we don't know the range of values \emph{a priori}, we don't know how to 
select $p$ to balance the two extremes. Therefore, we prefer the amount of 
rounding to be proportional to the size of the value itself (right column).

\begin{table}[h]\small
\centering
\begin{tabular}{c|c}
regular rounding & CAMP's rounding\\
\hline
10110\textbf{1011} $\rightarrow$ 10110\textbf{0000}
&\!\underline{1011}\textbf{01011} $\rightarrow$ 1011\textbf{00000}\\
00101\textbf{0011} $\rightarrow$ 00101\textbf{0000}
&\!00\underline{1010}\textbf{011} $\rightarrow$ 00101\textbf{0000}\\
00000\textbf{1010} $\rightarrow$ 00000\textbf{0000}
&\!00000\underline{1010} $\rightarrow$ 00000{1010}\\
00000\textbf{0111} $\rightarrow$ 00000\textbf{0000}
&\!000000\underline{111} $\rightarrow$ 000000{111}\\
\end{tabular}
\caption{Rounding with (binary) precision 4}\label{table:rounding}
\end{table}

The following proposition gives a bound on the number of distinct rounded values
for the cost-to-size ratio which in turn is an upper bound on the number queues
maintained by CAMP:

\begin{proposition}
If the original values for the cost-to-size ratio are integers in the range $1,\ldots,U$,
then the number of distinct rounded values is 
at most $(\lceil\log_2(U+1)\rceil-p+1) 2^p$
where $p$ is the selected precision.
\end{proposition}

{\noindent{\bf Proof:}}
 Let $x$ denote the value to be rounded. Since $x$ is in the range 1 through $U$, the binary representation of $x$ uses at most $\lceil\log_2(U+1)\rceil$ bits. Bit locations are numbered 1 through  $\lceil\log_2(U+1)\rceil$ with 1 being the lowest order bit. Let $b$ be maximum of $p$ and the location of the highest order non-zero bit in the binary representation of $x$. The scheme zeroes out all bits except those in locations $b, b-1,\dots,b-p+1$. There are at most $\lceil\log_2(U+1)\rceil-p+1$ possible values for $b$. For each value of $b$, there are $2^p$ possible rounded values encoded in bits $b, b-1,..,b-p+1$. Thus, the total number of distinct rounded values is $(\lceil\log_2(U+1)\rceil - p+1)2^p$.
{\hspace*{\fill}\rule{6pt}{6pt}\bigskip}

The competitive ratio of GDS is $k$ which means that on every sequence of requests, the overall cost of
GDS is within a factor of $k$ of the optimal algorithm that knows the entire request sequence in advance.
The proposition below shows that CAMP with precision $p$ approximates the behavior of GDS by a factor of 
$1+\epsilon $  where $\epsilon = 2^{-p+1}$ in the sense that CAMP obtains a competitive ratio of $(1+\epsilon)k$.
Thus, for sufficiently small $\epsilon$, the data structure would always evict the key-value pair with 
the true minimum priority.

\begin{proposition}
The competitive ratio of CAMP is $(1 + \epsilon)k$, where   $\epsilon = 2^{-p+1}$.
\end{proposition}

{\noindent{\bf Proof:}}
Consider an unrounded integer $x$ and denote its rounded value by $\bar{x}$.
We know that $\bar{x} \le x$ because  rounding only involves changing $1$'s to $0$'s.
Let $b$ be the location of the highest order bit in $x$. Then $\bar{x} \ge 2^{b-1}$.
Bits $1$ through $b-p$ are set to zero when $x$ is rounded. In the worst case,
the cleared bits are all 1. The amount that is subtracted from $x$ to get $\bar{x}$
is at most $2^{b-p}$.
Therefore, $(x - \bar{x})/\bar{x} \le 2^{b-p}/2^{b-1} = 2^{-p+1}$
and $x \le (1 + \epsilon)\bar{x}$, where $\epsilon = 2^{-p+1}$.

Now let $\sigma$ be a sequence of requests and let $\bar{\sigma}$ be the same request
sequence but with rounded cost-to-size ratios. Define $CAMP(\sigma)$ to be the
total cost of $CAMP$ on input $\sigma$ and let $OPT(\sigma)$ be the 
total cost of the optimal offline algorithm on input $\sigma$. 
CAMP makes the same eviction decisions on $\sigma$ as it does on $\bar{\sigma}$
because it rounds the cost-to-size ratios in $\sigma$. However, it pays potentitally a factor of
$(1+\epsilon)$ more
on each cache miss. Therefore $CAMP(\sigma)/(1 + \epsilon) \le CAMP(\bar{\sigma})$.
CAMP makes the same decisions as GDS on $\bar{\sigma}$
because the values are already rounded,
so $CAMP(\bar{\sigma}) = GDS(\bar{\sigma})$.
Since we know that GDS is $k$-competitive,
$GDS(\bar{\sigma}) \le k OPT(\bar{\sigma})$. Finally, OPT will have at least as large an overall
cost on $\sigma$ as it will on $\bar{\sigma}$ because all the cost-to-size ratios are
at least as large which means that $OPT(\bar{\sigma}) \le k OPT(\sigma)$. 
Putting it all together, we get
$$\frac{CAMP(\sigma)}{1+\epsilon} \le CAMP(\bar{\sigma}) = GDS(\bar{\sigma})  \le  k \cdot OPT(\bar{\sigma})  \le k \cdot OPT({\sigma}),$$
and CAMP is $(1 + \epsilon)k$-competitive.
{\hspace*{\fill}\rule{6pt}{6pt}\bigskip}

To use the integer rounding scheme we have described, we must first convert the cost-to-size 
ratio from a fraction to an integer. We cannot simply round since we may lose 
information regarding the relative order of magnitude among values that are 
less than 1. We can solve this problem by first dividing the cost-to-size ratio by a 
lower bound estimate on the smallest cost-to-size ratio that can ever occur. Then we perform the  
actual rounding to the nearest integer. The cost of each key-value pair is 
a non-negative integer, so 1 divided by the maximum size of any key-value 
pair can serve as a lower bound for the cost-to-size ratio. Thus, we are effectively
multiplying each cost-to-size ratio by the size of the largest key-value pair.
Although we do 
not know the maximum size of a key-value pair \emph{a priori}, we can 
determine it adaptively.  (The next paragraph describes why we do not simply use a large number such as the cache size.) A variable is used to hold the current maximum size 
observed so far. The variable is updated as soon as a referenced key-value 
pair is larger than the current maximum. For the sake of efficiency, we do 
not update the rounded priorities of all the key-value pairs in the KVS when 
a new lower bound on the cost-to-size ratio is determined. However, the new 
value is used for all future rounding. 

The rounding scheme makes no a priori assumptions as to the values of the cost-to-size ratios other than assuming that the ratio of the smallest to largest cost/size ratio is bounded by $U$. Converting all values to an integer is just a mathematically convenient way of expressing that assumption. The goal is to use the range 1 to $U$ as effectively as possible in expressing the range of cost-to-size ratios. The larger the value of $U$, the more space that must be set aside for potential LRU queues. The conversion from fractional values to integers is
achieved by multiplying all values by a fixed multiplier and
rounding to the nearest integer. Selecting a large number for the  multiplier would result in large rounded values
and would require a large upper bound $U$. This explains why we do not simply use the cache size for the multiplier.

In short, CAMP computes the $H$ value of a key-value pair in three steps. As 
the initial step, it converts the cost-to-size ratio to an integer. It then 
rounds the result by the pre-specified precision to an approximate value $c$. 
Finally, it assigns the key-value pair to the LRU queue associated with the 
value $c$. The key-value pair is assigned an $H$ value of $c + L$, where $L$ 
is the offset parameter used by GDS. 

%If $U$ is the largest integer resulting from the initial step in which the cost-to-size 
%ratio is converted to an integer, then the maximum number of LRU queues with 
%CAMP is $(\log_2(U)-p) 2^p$, where $p$ is the selected precision, which 
%satisfies\footnote{The explanation for this is as follows.  Let $X$ denote the value to be rounded. Since $X$ is in the range 1 through $U$, the binary representation of $X$ uses at most $\log_2(U)$ bits. Bit locations are numbered 1 through  $\log_2(U)$ with 1 being the lowest order bit. Let $b$ be maximum of $p$ and the location of the highest order non-zero bit in the binary representation of $X$. The scheme zeroes out all bits except those in locations $b, b-1,\dots,b-p+1$. There are at most $\log_2(U)-p$ possible values for $b$. For each value of $b$, there are $2^p$ possible rounded values encoded in bits $b, b-1,..,b-p+1$. Thus, the total number of distinct rounded values is $(\log_2(U) - p)2^p$.}
%$p < \lceil\log_2(U + 1)\rceil$.

\begin{figure}
    \centering
    \includegraphics[width=0.48\textwidth]{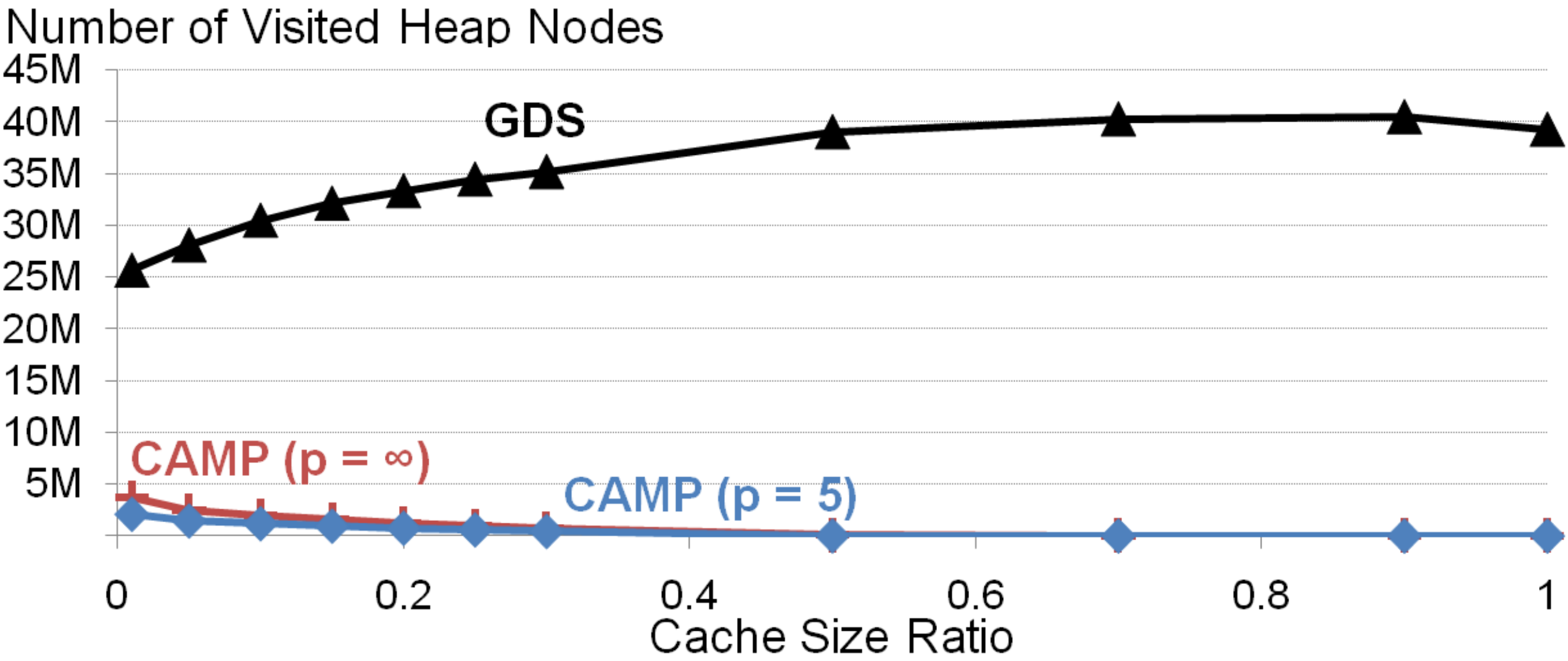}
    \caption{Number of visited heap nodes as a function of the cache size ratio.}
    \vspace{1.5em}
    \label{fig:nodeupdates}
\end{figure}
    
Figure~\ref{fig:nodeupdates} compares the number of visited heap nodes in a 
heap-based implementation of GDS and in CAMP when run using the trace-driven 
simulation of Section~\ref{sec:eval}. This quantity is an indication of the 
amount of runtime overhead of each implementation: in the case of GDS, this is 
the number of nodes that are visited when the heap is updated due to an 
insertion or deletion. In the case of CAMP, insertions and deletions from the 
queue comprise a constant time update to an LRU queue as well as the 
occasional update to the heap when the head of an LRU queue changes. 

There are two contributing factors to CAMP's significantly smaller number of 
node visits. First, the number of nodes in GDS's heap is equal to the number 
of key-value pairs in cache whereas the number of nodes in CAMP's heap is  
equal to the number of non-empty LRU queues, which is very small as noted in 
Figure~\ref{fig:precisionbuckets}. Since the number of node visits required by a heap
update grows logarithmically with the number of nodes in the heap, GDS's heap
updates can take longer than CAMP's. The second contributing factor is that 
GDS makes more heap updates than CAMP does. In particular, GDS updates its 
heap every time the priority of a key-value pair is updated. On the other
hand, CAMP only does so whenever the priority value of the head node of an LRU 
queue changes, or when an LRU queue is created or deleted.

Figure~\ref{fig:nodeupdates} shows the number of visited nodes by GDS 
increases as a function of the memory size.  This trend is reversed with 
CAMP.  The GDS curve increases because the number of nodes in the heap is 
equal to the number of items in the KVS and there are many more data items 
with a larger memory size. In contrast, the CAMP curve decreases because the 
number of heap 
nodes, which is equal to the number of non-empty LRU queues,  remains constant 
as a function of cache size. But since more items can be stored when the cache 
size increases, there are fewer updates to the cache. Hence, the decreasing 
curve.

\section{Evaluation}\label{sec:eval}

%\begin{table*}
%\begin{tabularx}{\textwidth}{|l|X|}
%\hline
%Miss rate         & Number of requests that observe a cache miss divided by number of issued requests. \\
%\hline
%Cost-miss ratio   & Sum of the costs of key-value references that observe a cache miss divided by the sum of the costs for all key-value references.\\
%\hline
%Cost              & Amount of time required to compute a key-value pair. \\
%\hline
%Cache Size Ratio  & Memory size divided by the total size of the key-value pairs competing for the memory.\\
%\hline
%\end{tabularx}
%\caption{Definitions of terms used.}
%\label{tbl:terms}
%\end{table*}

We used a social networking benchmark named BG~\cite{sumita13,bgycikm13,wbdb13,tpctc14} to generate 
traces of key-value references from a cache augmented database management 
system~\cite{jason13}.  BG emulates members of a social networking site 
viewing one another's profile, listing their friends, and other interactive 
actions.  The benchmark is configured to reference keys using a skewed pattern 
of access with approximately 70\% of requests referencing 20\% of keys.  A 
trace file consists of approximately 4 million rows.  Each row identifies a 
referenced key-value pair, its size, and cost.  Cost is either the time 
required to compute the key-value pair by issuing queries to the RDBMS or a 
synthetic value selected from \{1, 100, 10K\}\footnote{The values \{1, 100, 10K\} were chosen to simulate widely varying costs between key-value pairs.}.  
With the latter, each 
key-value pair is assigned one of the three possible values with equal 
probability.  Once a cost is assigned to a key-value pair, it remains in 
effect for the entire trace.

We implemented a simulator that consists of a KVS and a request generator to 
read a trace file and issue requests to the KVS.  The KVS manages a fixed-size 
memory that implements either the LRU or the CAMP algorithm.  Every time 
the request generator references a key and the KVS reports a miss for its 
value, the request generator inserts the missing key-value pair in the KVS.  
This results in evictions when the size of the incoming key-value pair is 
larger than the available free space.

The simulator quantifies two key metrics:  miss rate and cost-miss ratio.  
\emph{Miss rate} is the total number of requests that result in a KVS miss 
divided by the total number of requests in the sequence.  
\emph{Cost-miss ratio} is obtained by summing the costs for each request that 
results in a KVS miss divided by the sum of the costs for all the requests. 
With both, the first request to a particular key-value pair in the trace 
(called a {\em cold} request) is  not counted because any algorithm will fault 
on such requests. Since CAMP is tuned to minimize the total cost of serving 
the requests in the sequence, the cost-miss ratio is the primary metric used 
to quantify performance.  In the following, we report these metrics as a 
function of either the precision used by the CAMP algorithm or the \emph{cache 
size ratio}. The latter is the size of the KVS 
memory divided by the total size of the unique objects in the trace file.

\begin{figure}
    \centering\small
    \begin{subfigure}{0.48\textwidth}
        \includegraphics[height=3cm]{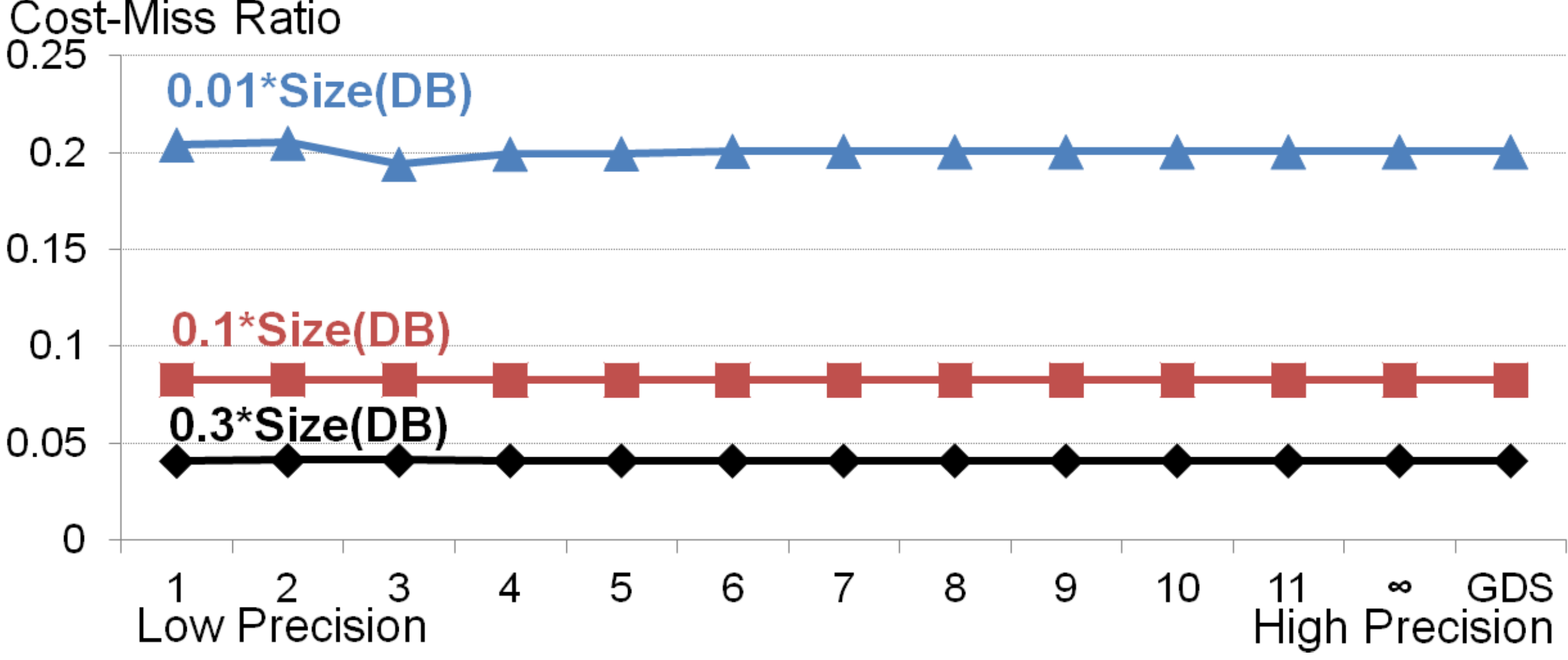}
        \caption{Cost-miss ratio as a function of precision.}
        \label{fig:precisioncost}
    \end{subfigure}\vspace{2em}
    \begin{subfigure}{0.48\textwidth}
        \includegraphics[height=3cm]{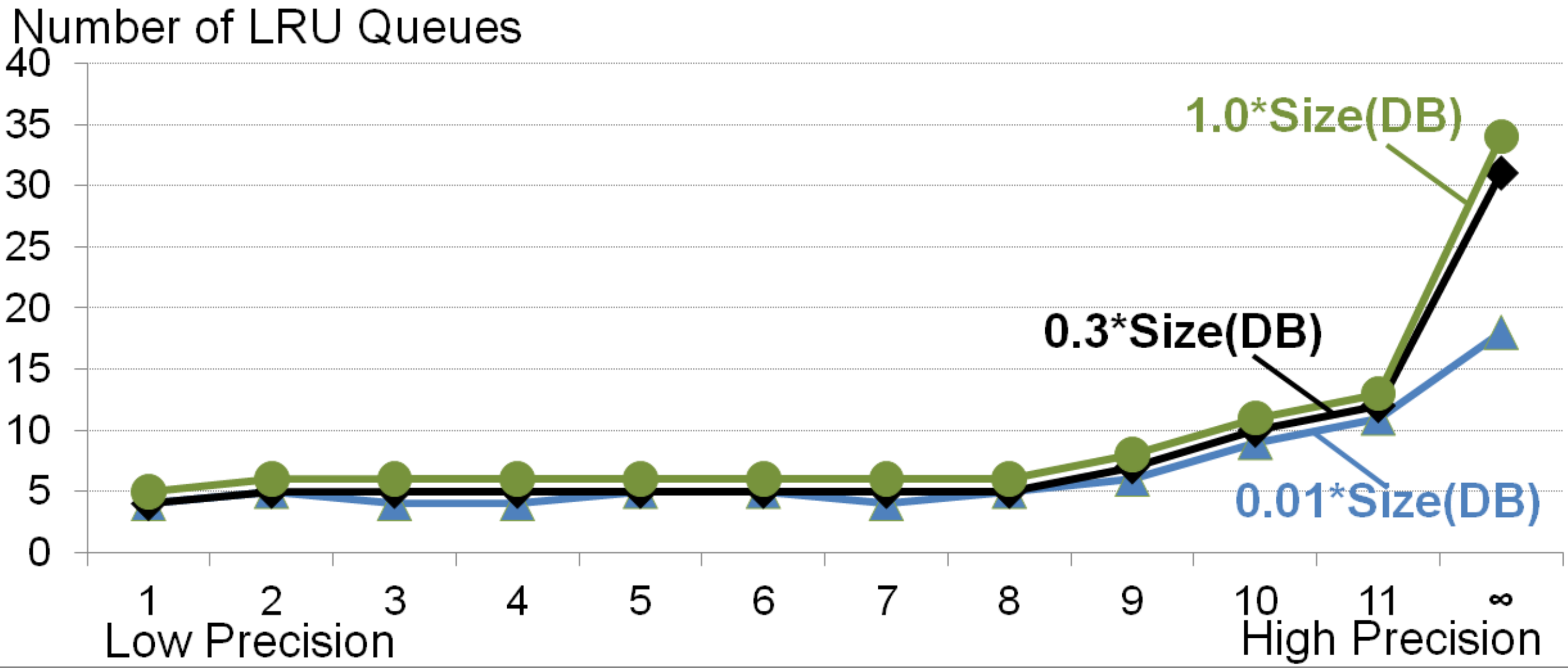}
        \caption{Number of LRU queues maintained by CAMP with different precisions.}
        \label{fig:precisionbuckets}
    \end{subfigure}\vspace{2em}
    \begin{subfigure}{0.48\textwidth}
        \includegraphics[height=3cm]{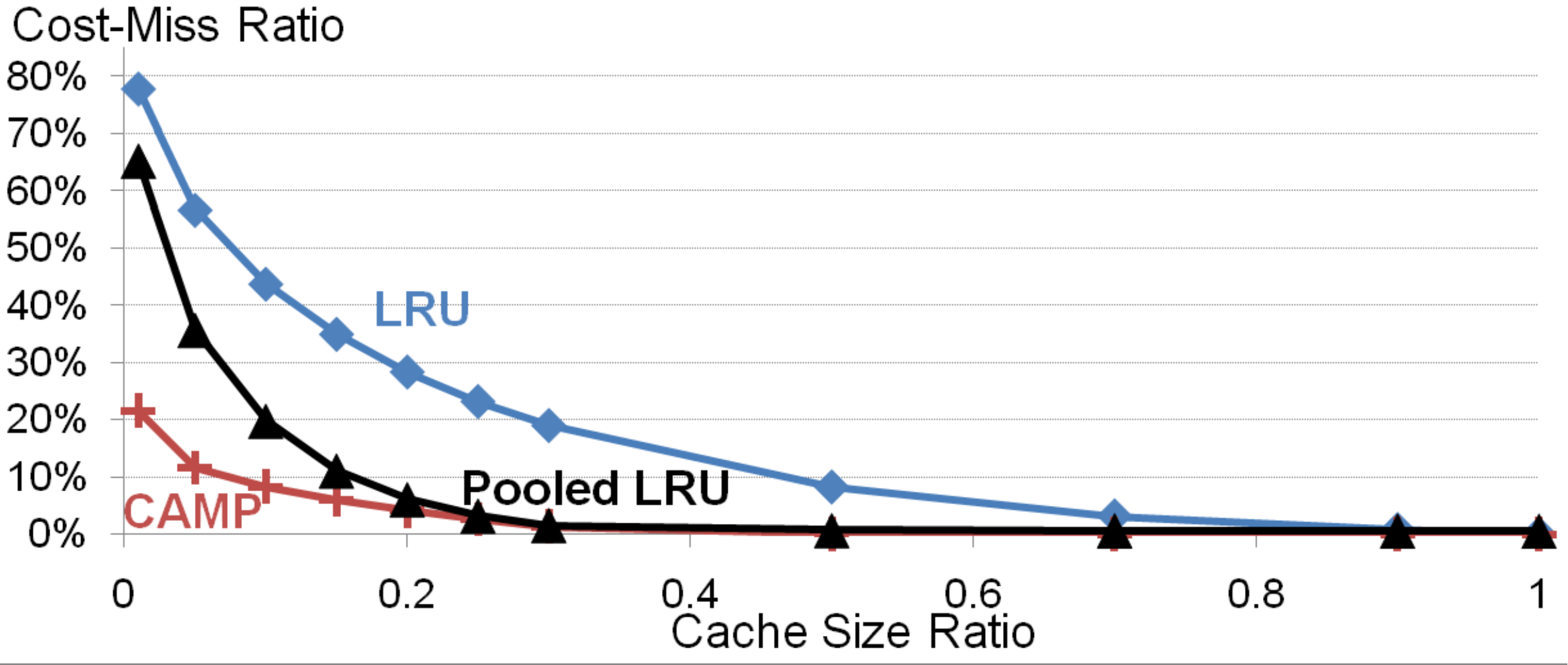}
        \caption{Cost-miss ratio as a function of the cache size ratio.}
        \label{fig:simcostratio}
    \end{subfigure}\vspace{2em}
    \begin{subfigure}{0.48\textwidth}
        \includegraphics[height=3cm]{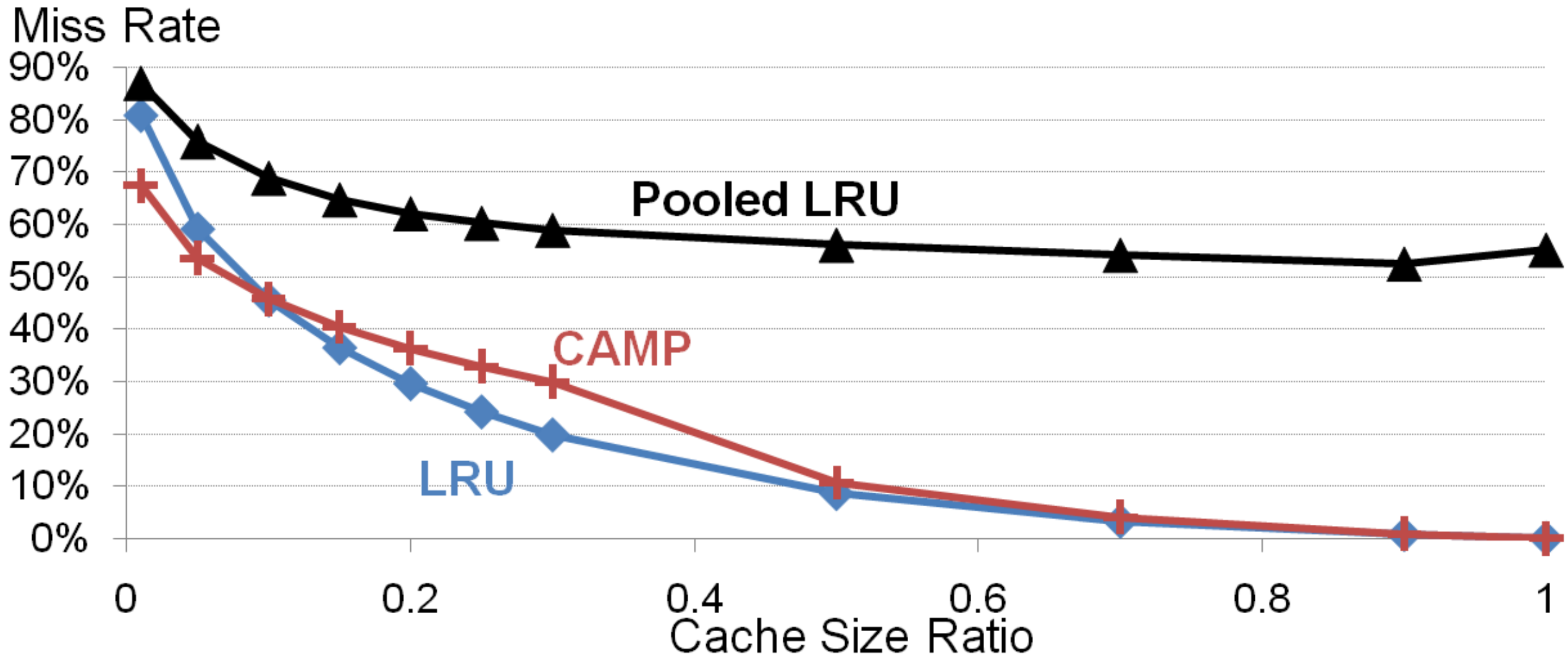}
        \caption{Miss rate as a function of the cache size ratio.}
        \label{fig:simmissratio}
     \end{subfigure}
     \caption{Simulation results with one trace and cost values selected from 1, 100, 10K.  With~\ref{fig:simcostratio} and~\ref{fig:simmissratio}, the precision of CAMP is set to 5.}\label{fig:simres}
\end{figure}

The precision of CAMP is a parameter that can be tuned for performance 
optimization. Small values of precision result in fewer LRU queues. With 
larger values of precision, replacement decisions are more finely tuned to 
differences in the cost-to-size ratio for each key-value pair. The graph in 
Figure~\ref{fig:precisioncost} shows the cost-miss ratio for CAMP as a 
function of the precision value. The three curves show the results for three 
different cache sizes. For the version labeled $\infty$, no rounding is done 
after the initial cost-to-size ratio is rounded to an integer. In other words, this version corresponds to the standard GDS algorithm.
Figure~\ref{fig:precisioncost} shows that there is almost no variation in 
cost-miss ratios for different precisions. More importantly, there is almost no difference between the cost-miss ratios of CAMP and standard GDS.

Figure~\ref{fig:precisionbuckets} shows the number of distinct LRU queues 
maintained by CAMP as a function of precision. The maximum number of queues 
possible is the number of distinct possible values for the cost-to-size ratio 
of the key-value pairs for a particular precision value. However, the KVS may 
not hold a key-value pair with a particular cost to size value, so at any 
given point in time, many of the queues are empty. 
Figure~\ref{fig:precisionbuckets} shows the actual number of non-empty queues 
at the end of the trace. Even for a very low level of precision, CAMP has at 
least five non-empty queues and outperforms LRU that has only one queue, see 
Figure~\ref{fig:simcostratio}.

In Figure~\ref{fig:simcostratio}, Pooled LRU is the partitioned-memory scheme 
described in~\cite{scalingmemcache}. This approach partitions the available 
memory into distinct pools.  Each pool employs LRU to manage its memory.  
Those key-value pairs with similar costs are grouped together according to 
their cost.  Different groups are assigned to different pools so that cheap 
and expensive key-value pairs do not compete with one another for the same 
memory.  This is not the same as CAMP, which adjusts the amount of memory 
used by each queue automatically, as demand fluctuates. 

To give Pooled LRU the greatest advantage, the amount of memory for each 
queue is computed in advance using the frequency of references to the 
different key-value pairs over the entire trace.  We experimented with 
different ways to partition the memory. In the first way, memory is allocated 
uniformly between the three queues. In the second way, the fraction of the 
total available memory assigned to each queue is proportional to the total 
cost of requests in the trace that belong to a particular pool.

With the BG benchmark-generated trace using synthetic cost values selected 
from \{1, 100, 10K\}, Pooled LRU constructs three pools. These pools have 
approximately the same number of key-value pairs, frequency and size, where 
the frequency of a pool is the number of references made to key-value pairs 
in that pool and the size is the amount of memory needed to store all key-value 
pairs in that pool. With a uniform partitioning of memory, Pooled LRU has
both a cost-miss ratio and miss rate similar to LRU. This is a consequence
of the key-value pairs assigned to each pool having the same  frequency and size. 
The performance is so close, that we only 
display the cost-miss ratio for LRU in Figure~\ref{fig:simcostratio}. When 
memory is partitioned using cost, Pooled LRU improves over LRU's cost-miss 
ratio. Furthermore, it is able to match CAMP's cost-miss ratio when given a 
large enough cache size by assigning practically all of it to the most 
expensive pool. However, this improvement is at the expense of a 
significantly worse miss rate (see Figure~\ref{fig:simmissratio}), since, 
even with a large cache size, the cheapest pool has nearly a 100\% miss rate 
and the second pool has a miss rate of 65\%.

\begin{figure}
    \centering\small
    \begin{subfigure}{0.48\textwidth}
        \includegraphics[width=\textwidth]{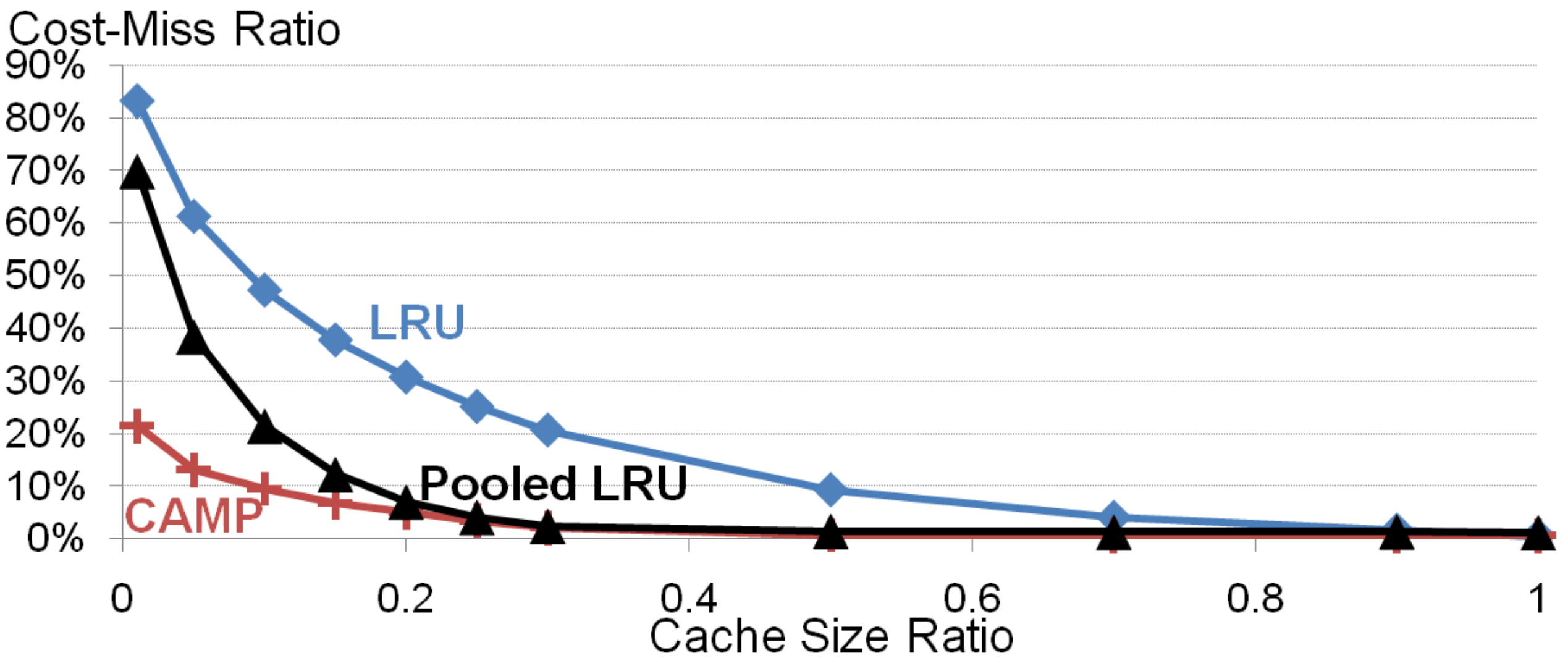}
        \caption{Cost-miss ratio as a function of cache size ratio.}
        \label{fig:calcification-cost}
    \end{subfigure}\vspace{2em}
    \begin{subfigure}{0.48\textwidth}
        \includegraphics[width=\textwidth]{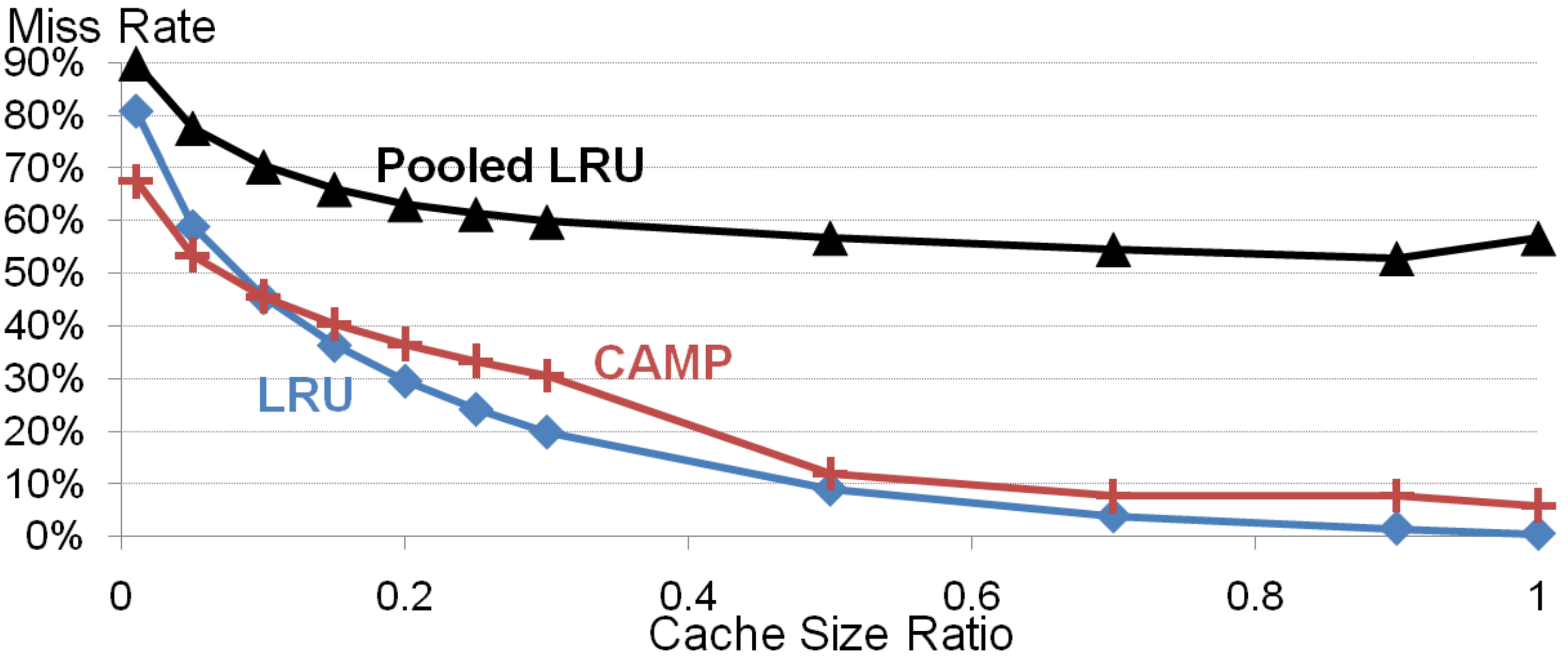}
        \caption{Miss rate as a function of cache size ratio.}
        \label{fig:calcification-miss}
    \end{subfigure}\vspace{2em}
    \begin{subfigure}{0.48\textwidth}
				\includegraphics[width=\textwidth]{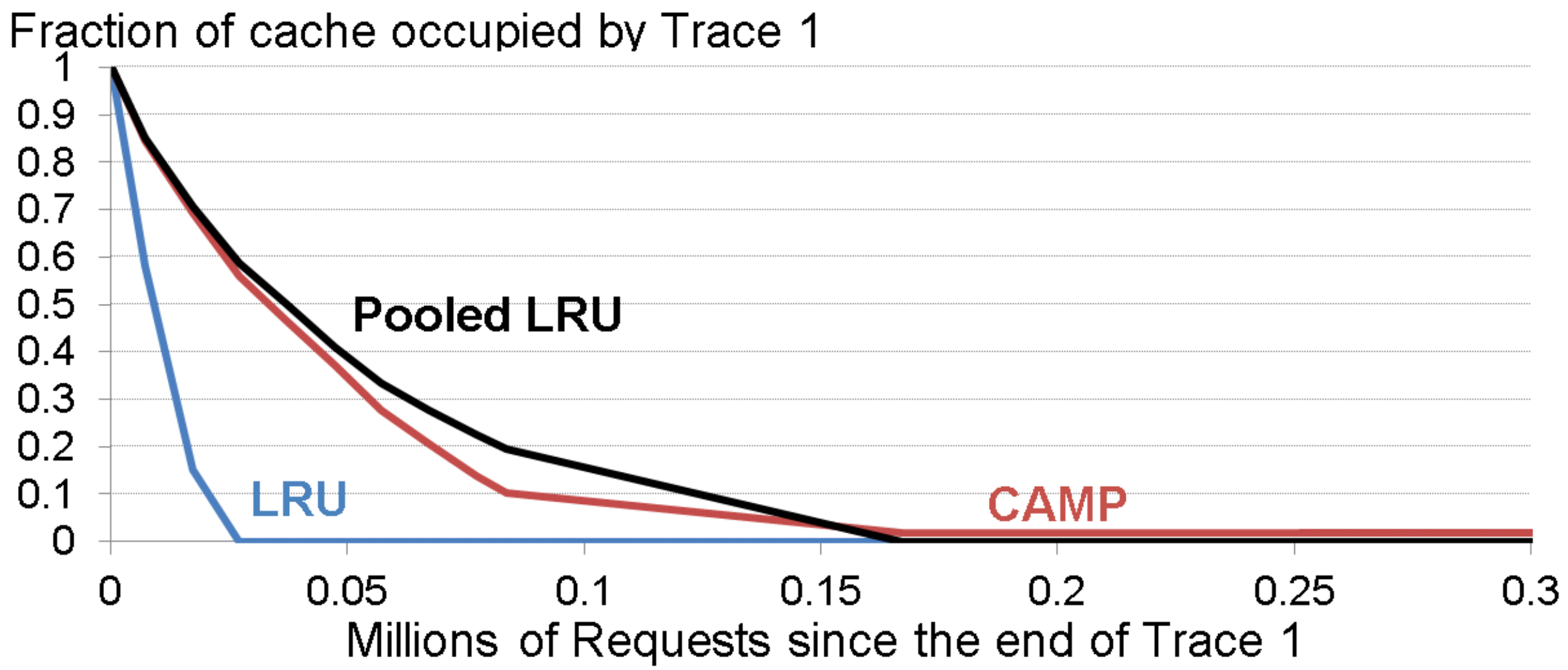}
        \caption{Fraction of cache occupied by trace 1 items, with cache size ratio set at 0.25.}
        \label{fig:calcification-time-0.25}
    \end{subfigure}\vspace{2em}
    \begin{subfigure}{0.48\textwidth}
				\includegraphics[width=\textwidth]{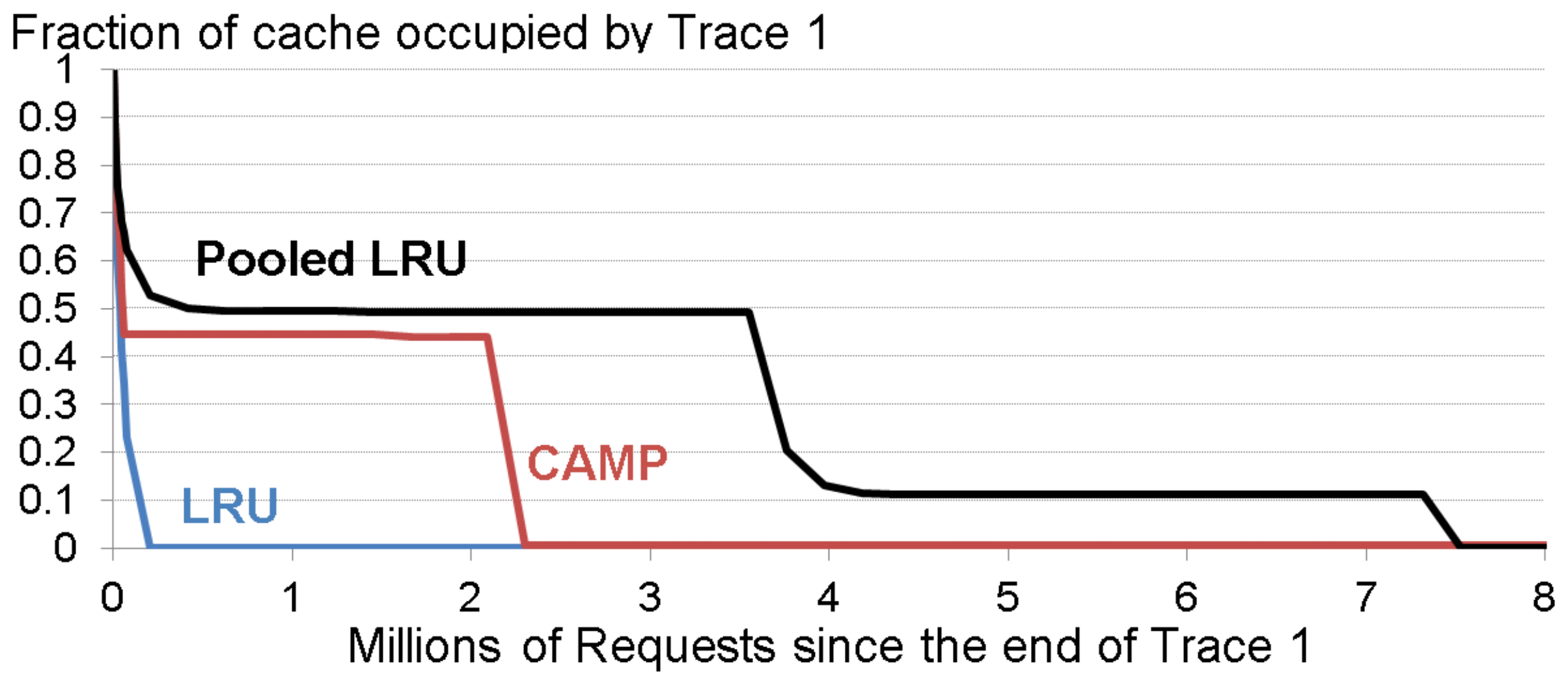}
        \caption{Fraction of cache occupied by trace 1 items, with cache size ratio set at 0.75.}
        \label{fig:calcification-time-0.75}
    \end{subfigure}
     \caption{Simulation results with changing access patterns.}\label{fig:calcification}
\end{figure}

\subsection{Evolving Access Patterns}\label{subsec:evolving}
A key feature of CAMP is its ability to adapt to evolving access patterns by evicting
those key-value pairs that were hot in some distant past.
This includes expensive key-value pairs.
%Below, we present results from an experiment that is adversarial to CAMP.
Obtained results show CAMP adapts effectively when compared to LRU and Pooled LRU.
Moreover, the overall cost-miss ratio and miss rate trends remain the same as the results of
Figure~\ref{fig:simres}.

%%In this experiment, we used ten different traces referencing unique key-value pairs
%%%to issues requests for the different algorithms.
%%Each trace file consists of 4 million key-value references and
%%identified as Trace File 1, TF1, Trace File 2, TF2, etc.
In this experiment, we used ten different traces back to back. Each trace file consists of 4 million key-value references. Moreover, requests from different traces are given distinct identification, so any request from a given trace file will never be requested again after that trace.
The  trace files are
identified as Trace File 1, TF1, Trace File 2, TF2, etc.
They are adversarial to CAMP because each trace file is generated using a skewed distribution
of access as described at the beginning of this section.
This means once the simulator switches from TF1 to TF2,
none of the objects referenced by TF1 are referenced again.
The same holds true for all other trace files, emulating a sudden shift in access patterns where
expensive objects that were referenced frequently are never referenced again.

We conducted the above experiment for a variety of cache sizes and observe CAMP adapts across
the different trace files to provide a cost-miss ratio and miss rate similar to those observed
in the previous section, see Figure~\ref{fig:calcification-cost} and~\ref{fig:calcification-miss}.
Figures~\ref{fig:calcification-time-0.25} and~\ref{fig:calcification-time-0.75} show the fraction of KVS memory occupied by the key-values of
TF1 for two different KVS memory size ratios, 0.25 and 0.75.
These two figures show how well the different techniques adapt to the sudden change in access
patterns.
The x-axis of these two figures it the number of key-value references issued relative to the
start of TF2 in millions of requests.
The transition to a different TF is at the 4 million tick mark of the x-axis.
The y-axis shows the fraction of KVS memory size occupied by key-value pairs of TF1.

With a small cache size (see Figure~\ref{fig:calcification-time-0.25} with a cache size ratio of 0.25),
all three algorithms evict key-value pairs referenced by TF1 quickly.
LRU is the quickest, evicting all key-value pairs of TF1 after 21,000 references of TF2.
It is followed by Pooled LRU with 131,000 references of TF2.
CAMP evicts most of TF1 key-value pairs quicker than Pooled LRU and slower than LRU.
It does not evict all TF1 key-value pairs (those with the highest cost-to-size ratio)
until 7.7 million references, close to the end of TF3.
However, these items occupy less than 2\% of the total cache size. 

With a larger cache size (see Figure~\ref{fig:calcification-time-0.75} with a cache size ratio of 0.75),
both CAMP and Pooled LRU behave in a step function with CAMP evicting a majority
of TF1 key-value pairs faster than Pooled LRU.
Once again, LRU is quickest as it considers recency of references.
Pooled LRU evicts all key-value pairs referenced by TF1 after 7.3 million requests
are issued, close to the end of TF3.
CAMP maintains a few of the most expensive key-value pairs of TF1 even
after 40 million requests are issued.
However, these occupy less than 0.6\% of the available KVS memory.

Let us analyze the behavior of each algorithm. LRU evicts the key-value pairs requested in TF1 when the total size of newer key-value pairs is greater than the cache size, which occurs before the transition to TF3 regardless of cache size. The jump in eviction time at cache size ratio 1 corresponds to the fact that the key-value pair that causes the total
size of requested key-value pairs to exceed the cache size is the first key-value pair requested in TF3.

The sudden eviction of large portions of TF1 key-value pairs by Pooled LRU at large cache sizes 
(see Figure \ref{fig:calcification-time-0.25})
correspond to the introduction of new key-value pairs at the beginning of TF3 and TF4, which occur at around the 4 millionth and 8 millionth mark. As in the first experiment, Pooled LRU pools key-value pairs by cost, and for a fixed cache size, each pool is allotted a portion of the cache proportional to the cost value. Since the only occurring cost values are 1, 100 and 10K, 99\% of the cache is dedicated to the pool of expensive key-value pairs. On the other hand, the expensive key-value pairs from a single TF only occupy a third of the maximum cache size, so that a cache size ratio of 2/3 can store all expensive key-values from two TFs, and a cache size ratio of 1 can store those of 3 separate TFs. Now the point at which all TF1 key-value pairs are evicted occurs when the total size of the expensive key-value pairs requested in a subsequent TF exceeds
the cache size. When the cache size ratio is 1/3 or less, this occurs before the end of TF2 (4 million requests). At a cache size ratio of 2/3 or higher, the eviction time occurs during TF4 (roughly between 8 and 12 million requests).

Finally, CAMP maintains an LRU queue for each cost-to-size ratio, and unlike Pooled LRU, these queues can be resized dynamically. Because key-value pairs requested at a later time can have higher priority of being evicted than those requested earlier, CAMP only guarantees that those requested after TF1 that have the highest cost-to-size ratio will have lower priority than any TF1 request. This observation yields the loose guarantee that all TF1 key-value pairs will be evicted by the time the total size of all newer requested items with the highest cost-to-size ratio reaches the cache size. According to Figure~\ref{fig:calcification-time-0.75}, for a cache size ratio of 0.75, there are cache-resident still key-value pairs from TF1 at the end of the simulation. This is explained by the fact that the highest cost-to-size key-values contribute less than 1/20th of the maximum cache size per TF. Together over the whole trace, these key-value pairs will fit in caches with cache size ratios greater than 0.5. Hence, the condition guaranteeing the eviction of all TF1 items is  
satisfied with the .25 cache size of Figure \ref{fig:calcification-time-0.25} and not satisfied with the .75 cache size of Figure \ref{fig:calcification-time-0.75}. 

\subsection{Other Traces}

\begin{figure}
    \centering\small
    \includegraphics[width=0.48\textwidth]{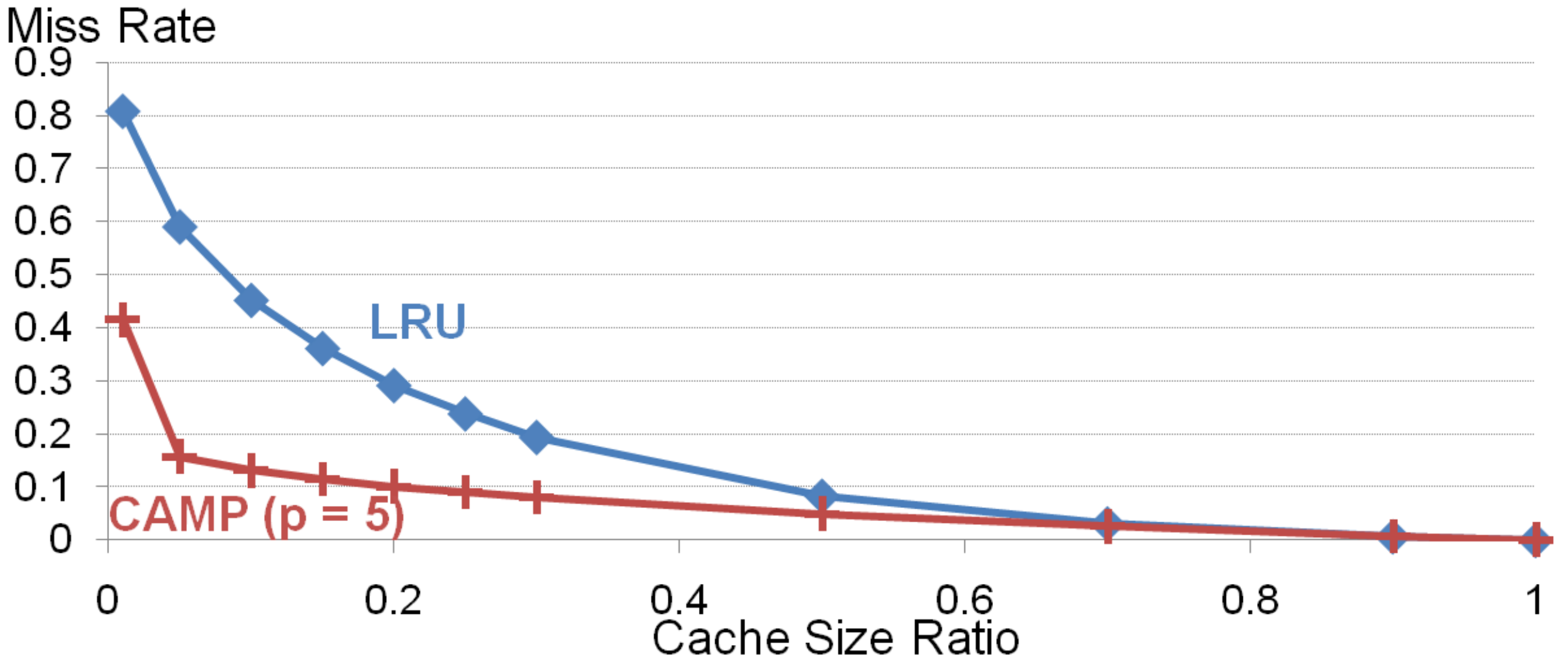}
    \caption{Miss rate as a function of cache size with variable sized key-value pairs and constant cost.}
    \label{fig:cost1-miss}
\end{figure}

\begin{figure}
    \centering\small
    \begin{subfigure}{0.48\textwidth}
        \includegraphics[width=\textwidth]{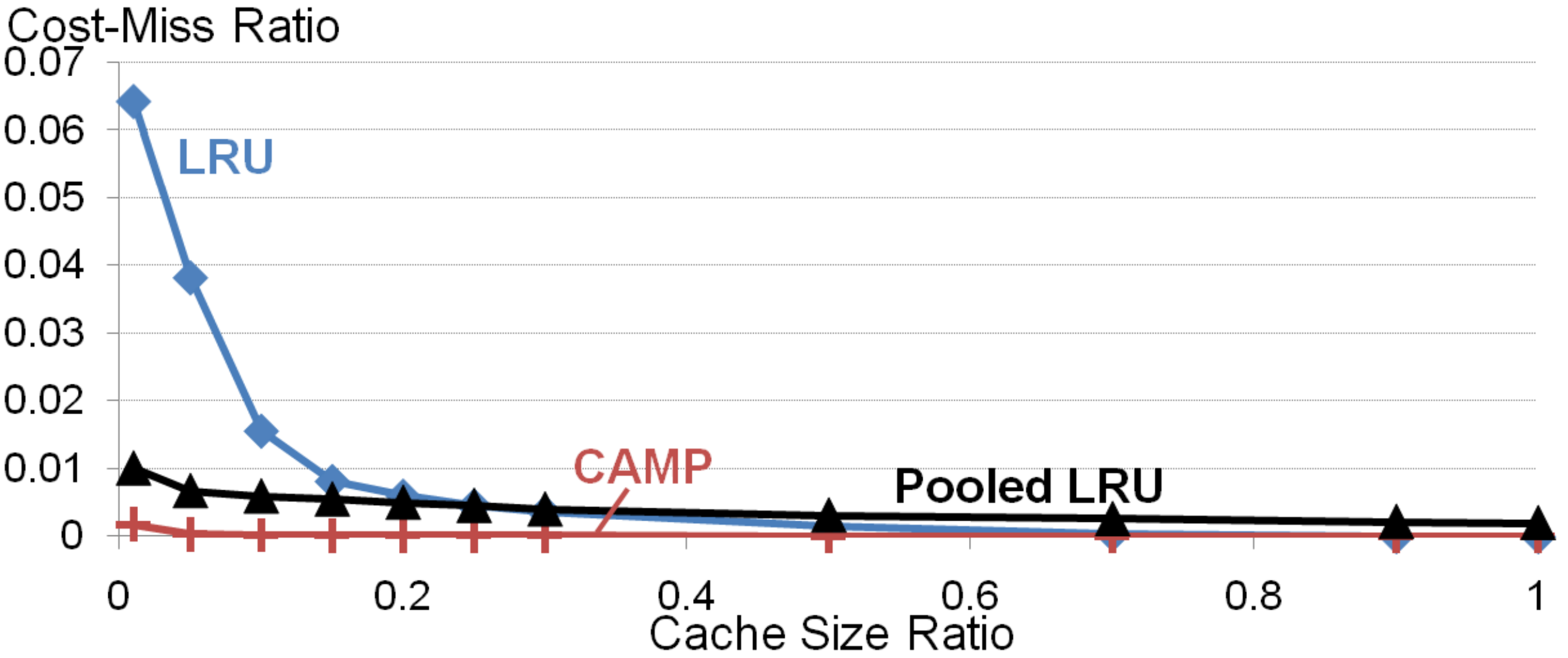}
        \caption{Cost-miss ratio as a function of cache size ratio.}
        \vspace{1.5em}
        \label{fig:bigtrace-cost}
    \end{subfigure}\\
    \begin{subfigure}{0.48\textwidth}
        \includegraphics[width=\textwidth]{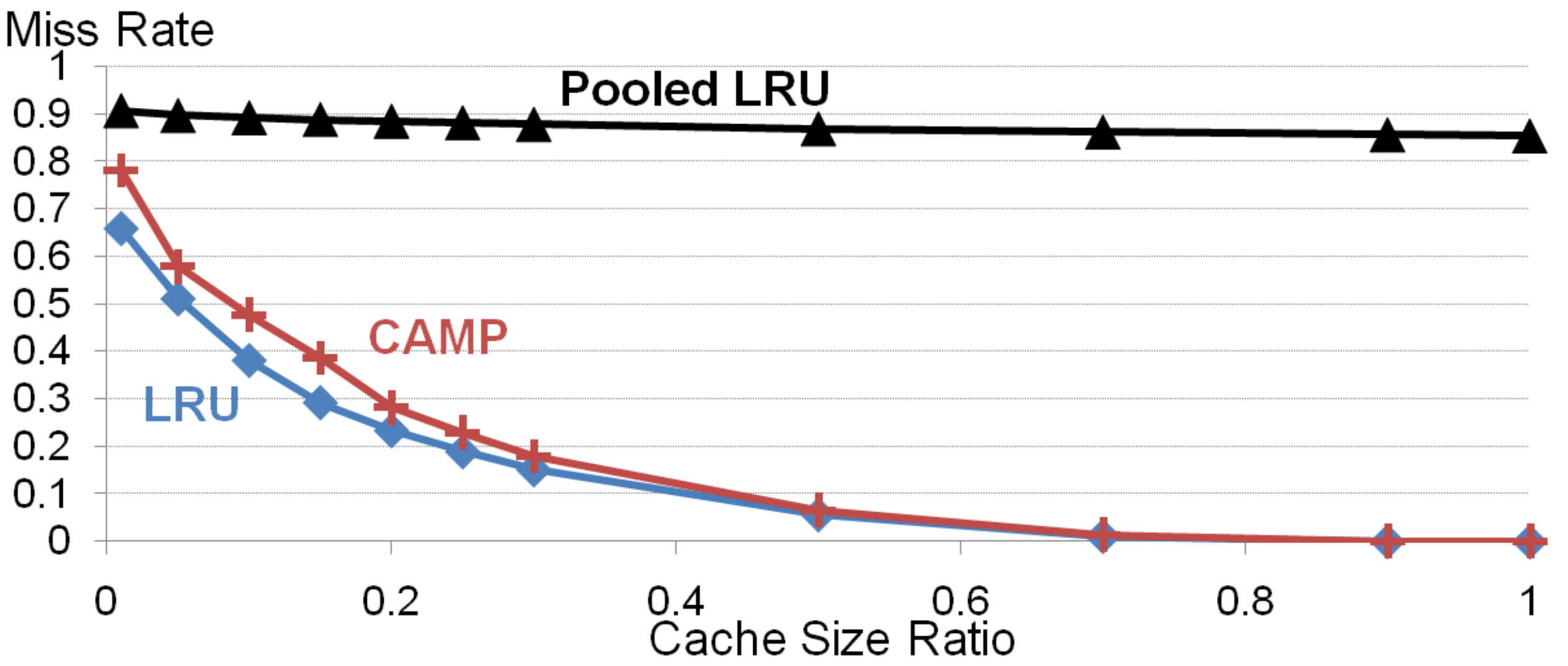}
        \caption{Miss rate as a function of cache size ratio.}
        \vspace{1.5em}
        \label{fig:bigtrace-miss}
    \end{subfigure}\\
    \begin{subfigure}{0.48\textwidth}
        \includegraphics[width=\textwidth]{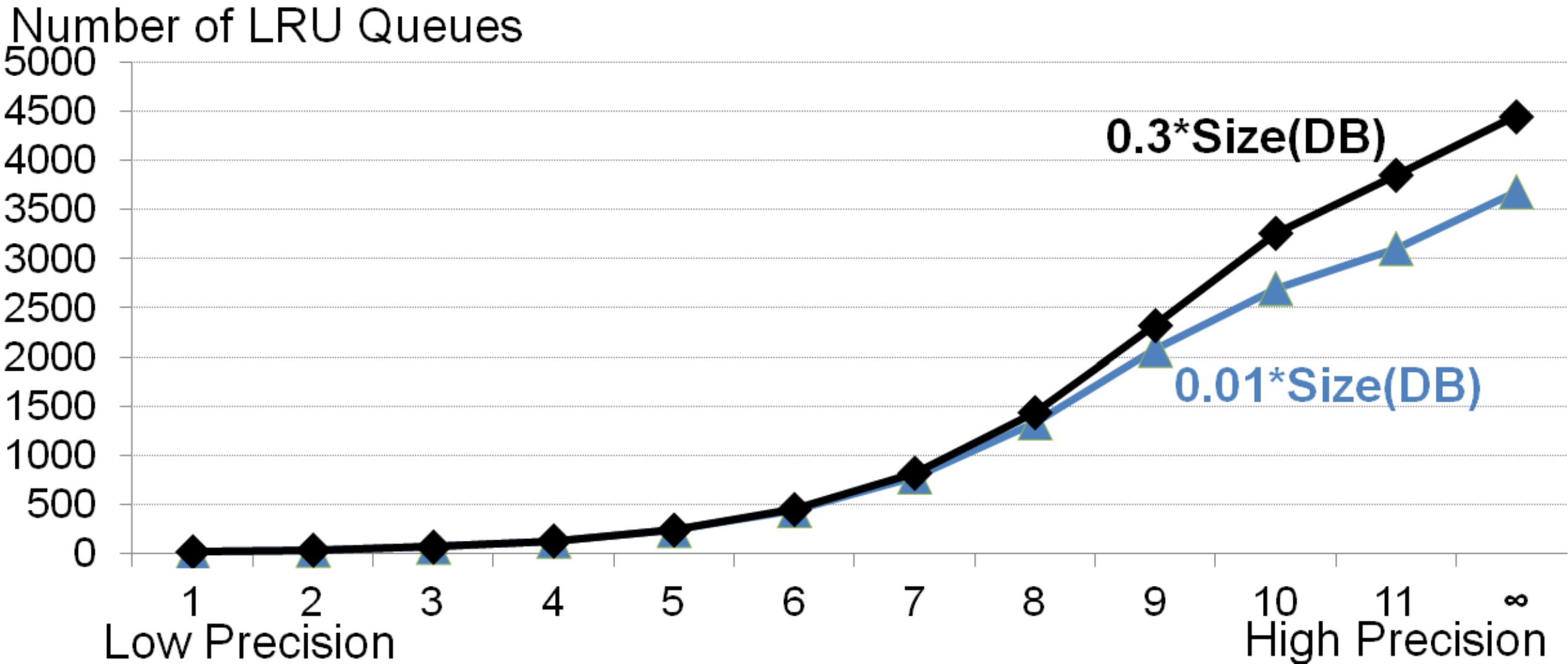}
        \caption{Number of queues as a function of precision.}
        \vspace{1.5em}
        \label{fig:bigtrace-queues}
    \end{subfigure}
     \caption{Simulation results with equi-sized key-value pairs and variable costs.}\label{fig:othertraces}
\end{figure}

The trends reported in Subsection~\ref{subsec:evolving} hold true with other traces.
The most insightful results are obtained with the two possible extremes,
namely, variable sized key-value pairs with almost similar costs and
equi-sized key-value pairs with varying costs.
We describe these in turn.

With variable sized key-value pairs whose cost is identical, CAMP renders
small key-value pairs KVS resident, providing a lower miss rate when
compared with LRU, see Figure~\ref{fig:cost1-miss}.
Pooled-LRU constructs one pool and behaves the same as LRU.
Because the cost of the key-value pairs is set to 1, 
the cost-miss ratio of a technique is equal to its miss rate.
Hence Figure~\ref{fig:cost1-miss} can also be interpreted as the cost-miss ratio of
LRU and CAMP as a function of cache size ratio.

With equi-sized key-value pairs that incur a different cost, CAMP continues
to provide a superior cost-miss ratio to both LRU and Pooled-LRU, see Figure~\ref{fig:bigtrace-cost}.
With a limited amount of memory, CAMP's miss rate is slightly worse than that of LRU as
it favors high cost key-value pairs, see Figure~\ref{fig:bigtrace-miss}.
With Pooled-LRU, we were challenged in determining the size of different pools
as there was no clear way to partition the range of different costs.
In the original trace where key-value pairs could only have one of three
different cost values, items were pooled by their cost value. Here, we opted
to pool items by range of cost values. Specifically, the ranges were 1 to 100,
100 to 10,000 and 10,000 and beyond. The available memory was then divided
among the three pools in such a way that each pool received an amount
proportional to the lowest cost value in its range.
With this assignment, Pooled-LRU results in a superior cost-miss ratio with small
cache-size ratios.
With larger cache size ratios, its partitioning of space makes it inferior to both
LRU and CAMP. 

In comparison to the trace with three distinct cost values (Figure~\ref{fig:precisionbuckets}),
the trace with equi-sized key-value pairs has key-value pairs with many more
distinct cost values. Therefore, for this trace, CAMP
creates a larger number of LRU queues when no rounding takes place, see
Figure~\ref{fig:bigtrace-queues}. This 
increase is due to the fact that there is a greater number of cost-to-size ratios
as a result of the considerably larger set of cost values. With additional
rounding, \emph{i.e.} at lower precision, the number of LRU queues in the two
traces decrease significantly and converge without any sizable performance
degradation.

\section{An Implementation}\label{sec:impl}
We implemented CAMP in the IQ Twemcache, a modified version of the Twemcache 
v2.5.3~\cite{twencache} that implements the IQ framework~\cite{iq14}.  This implementation 
computes the cost of a key-value pair by noting the timestamp of a miss 
observed by a get (iqget) and the subsequent insertion of the computed value 
using a set (iqset).  The difference between these two \mbox{timestamps} is 
used as the cost of the key-value pair.

The approach taken to provide recomputation time is to use the service time to compute a key-value pair (and piggybacked as a part of the KVS put).  Application provided hints are another possibility.  

CAMP does not need to address the issue of malicious applications with misleading costs, because it is intended for use in a middleware deployed in a trusted environment (data center) with no adversary.

\begin{figure}
    \centering
    \begin{subfigure}{0.48\textwidth}
        \includegraphics[width=\textwidth]{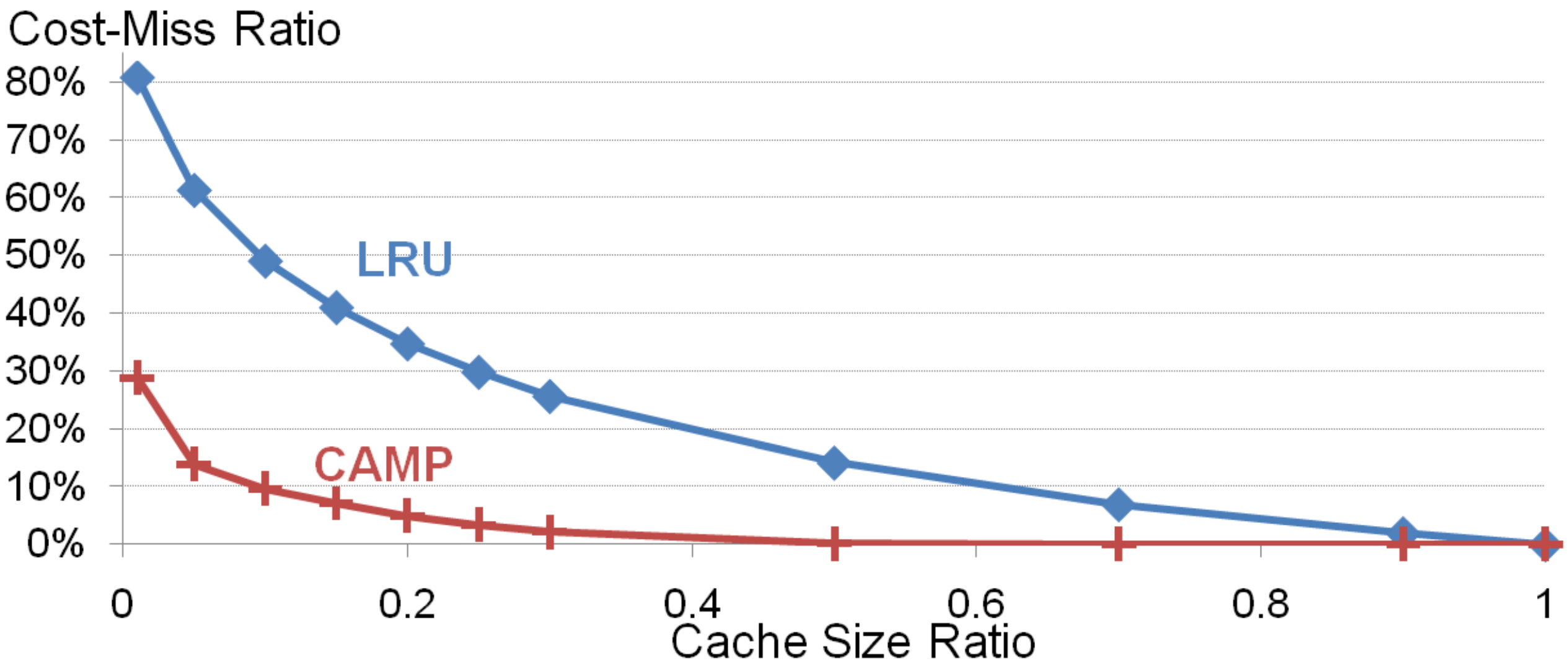}
        \caption{Cost-miss ratio as a function of the cache size ratio.}
        \vspace{1.5em}
        \label{fig:implres-costratio}
    \end{subfigure}
    \hfill
    \begin{subfigure}{0.48\textwidth}
        \includegraphics[width=\textwidth]{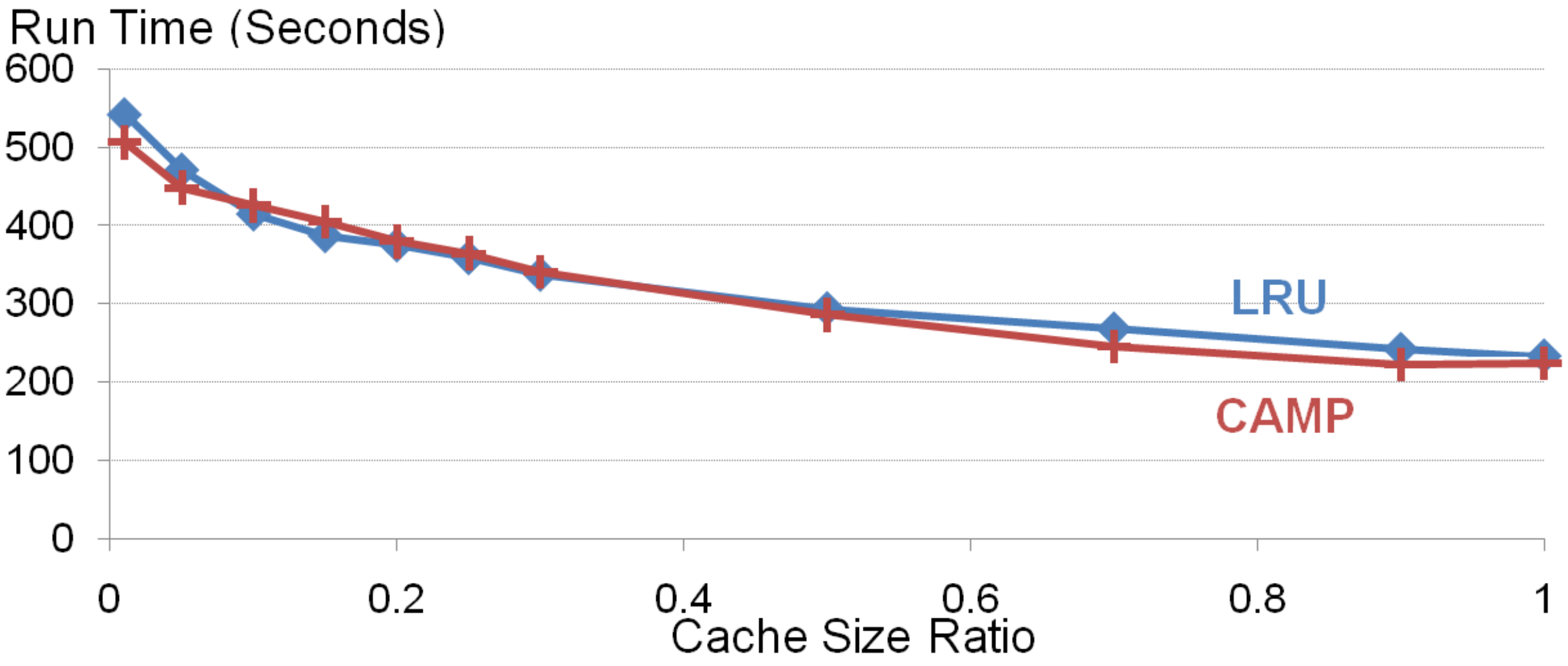}
        \caption{Run time as a function of the cache size ratio.}
        \vspace{1.5em}
        \label{fig:implres-runtime}
    \end{subfigure}
    \hfill
    \begin{subfigure}{0.48\textwidth}
        \includegraphics[width=\textwidth]{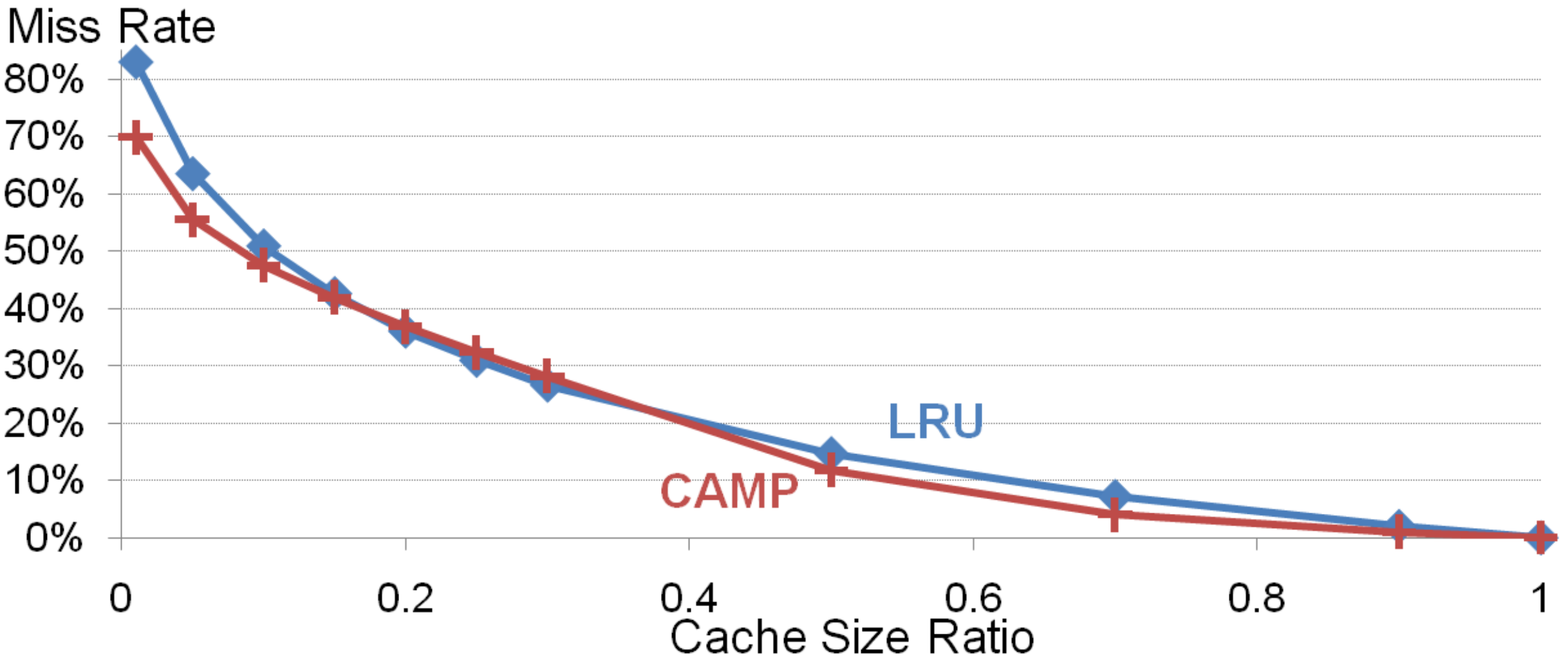}
        \caption{Miss rate as a function of the cache size ratio.}
        \label{fig:implres-missratio}
    \end{subfigure}
    \caption{Implementation results, where the precision of CAMP is set to 5.}
    \label{fig:implres}
\end{figure}

We developed an application that implements the request generator of 
Section~\ref{sec:eval} by reading a trace file and issuing requests to the KVS 
using Whalin client version 2.6.1~\cite{whalin2}.  We used the trace file with 
synthetic costs of \{1, 100, 10K\} per discussions of 
Section~\ref{sec:eval}.  Figure~\ref{fig:implres-costratio} shows the observed 
cost-miss ratio with LRU and CAMP as a function of different cache size 
ratios.  CAMP incurs a significantly lower cost for the missing keys with 
smaller cache sizes.  This difference becomes smaller with larger cache sizes 
because the KVS miss rate drops.  These results are consistent with those 
reported in Section~\ref{sec:eval}.

%\begin{figure}
%\begin{center}
%\psfig{figure = images/runtime.eps, width=0.3\textwidth}
%\caption{Run time as a function of the cache size ratio.}\label{fig:runtimesizeimpl}
%\end{center}
%\end{figure}

Figure~\ref{fig:implres-runtime} shows the amount of time required to run the 
trace with both LRU and CAMP.  It includes the following:  (1) the time for 
either LRU or CAMP to process a cache hit and to make replacement decisions 
when the memory is exhausted; (2) the time to transmit a key-value pair across 
the network (with both a cache hit and a cache insert); and (3) the time to 
copy a key-value pair across the different software layers.  The results show 
that CAMP provides response times comparable to those of LRU. If the cost was 
included in the reported response times, CAMP would have been significantly 
faster than LRU, resembling the results reported in Section~\ref{sec:eval}.  
Here, we wanted to show results that demonstrate that an implementation of 
CAMP is as fast as LRU while ignoring the cost associated with keys. 

%\begin{figure}
%\begin{center}
%\psfig{figure = images/missratio.eps, width=0.3\textwidth}
%\caption{Miss rate as a function of the cache size ratio.}\label{fig:missratesizeimpl}
%\end{center}
%\end{figure}

With both CAMP and LRU, the run time decreases as a function of cache size.  
The explanation for this is as follows.  
A get($k_i$) that observes a miss is followed by a set($k_i$,$v_i$).  With a 
small cache size that is full, the set operation must evict one or more 
existing key-value pairs and write the new $k_i$-$v_i$ in the cache.  This 
write operation requires copying of the $k_i$-$v_i$ from the network buffer 
into the memory of the cache manager.  A larger cache size reduces the number 
of such operations because a higher percentage of get($k_i$) operations 
observe a cache hit, see Figure~\ref{fig:implres-missratio}.  This explains 
why the run time improves with both LRU and CAMP using a larger cache size.  
This also explains why CAMP provides a faster response time than LRU for those 
cache sizes that provide a lower miss rate, \emph{i.e.}, cache ratio of 0.01.

\subsection{Discussion}

%Scalability can be achieved with the presence of multiple LRU queues, which enables several concurrent threads to access the different LRU queues independently.  Each LRU queue is represented as an array of queues with the keys hash partitioned across its array.  This provides for vertical scaling by allowing many concurrent threads accessing the same logical LRU queue to manipulate different physical LRU queues independently.  A larger number of physical LRU queues per logical LRU queue may lower the quality of the eviction decisions.  A key strength of CAMP is that its heap data structure (which requires synchronization) is updated only when the head of a logical LRU queue changes.

CAMP is suitable for use with multi-core processors as it minimizes the delay
incurred with concurrent threads racing with one another to read and write
CAMP's data structures.  Several features of CAMP enable it scale vertically.
First, it only updates its heap data structure (which requires synchronized access)
when the head of a LRU queue changes value instead of per eviction. 
Second, different threads may update different LRU queues simultaneously
without waiting for one another.
Finally, CAMP may represent each LRU queue as multiple physical
queues and hash partition keys across these physical queues
to further enhance concurrent access.

CAMP can be extended with use with a hierarchical cache (using SSD, disk, or both) which may persist costly data items.  With any finite cache size, should the data set size exceed the cache size, an algorithm must make an eviction decision.  CAMP systematically renders such decisions by considering size and cost of key-value pairs, and recency of references with a two level cache.

\section{Related Work}\label{sec:related}

Management of KVS memory space must address two key questions:
First, how should memory be assigned to a key-value pair?
Second, what key-value pairs should occupy the available memory?
In the context of online algorithms that manage memory, the second
question must identify what key-value pair should be evicted
when there is insufficient space for an incoming
key-value pair.
CAMP provides an answer to this question.
Below, we survey the state of the art and describe how CAMP is different.

LRU-K~\cite{oneil93}, 2Q~\cite{2Q}, and ARC~\cite{arc03} are adaptive replacement
techniques that balance between the recency and the frequency features of
a workload continuously, improving cache hit rate for fix sized disk
pages with a constant cost.
CAMP is different in that it considers both the size of a key-value
pair and its cost.
CAMP is an efficient implementation of the GDS algorithm~\cite{irani97},
visiting a significantly fewer number of heap nodes than GDS,
see the discussion of Figure~\ref{fig:nodeupdates} in Section~\ref{sec:camp}.

Similar to CAMP, GD-Wheel~\cite{li13} strives to enhance the efficiency of GDS. There are, however, some significant differences in approach. GD-Wheel rounds the overall priority for each key-value pair instead of the cost-to-size ratio which makes it difficult to evaluate the cost of approximation. The GD-Wheel study does not give a direct comparison between GD-Wheel and GDS. With CAMP we are able to give well-defined guarantees on the competitive ratio of CAMP relative to GDS. Moreover, GD-Wheel does not address how to select their precision parameter $N$ or give an empirical characterization of performance as a function of precision. Finally, GD-Wheel must implement occasional migration procedures wherein all the key-value stores within a GD-Wheel are migrated to the next level. CAMP does not require such a migration step as it uses the cost-to-size ratio as the basis of the rounding scheme (which does not change while a key-value pair is in the cache).

While deciding how space should be assigned (answer to the first question)
is a different topic, there are implementations that strive to answer this
question in combination with a replacement technique.
For example, the memory used by a Twemcache server instance is managed
internally using a slab allocation system~\cite{Bonwick94} in combination with LRU.
Below, we describe this technique and how it is different than CAMP.

Twemcache divides memory into fix
sized slabs (chunks of memory), the default size being 1 Megabyte. 
Each slab is then assigned a slab class and further sub-divided 
into smaller chunks based on its slab class. For example, a slab class of 1,
the smallest size, would have a chunk size of 120 bytes. 
This means that a single slab of class 1 can fit 8737 (1 MB / 120 byte) chunks.
Every subsequent higher slab class uses chunk sizes that are approximately a 
factor of 1.25 larger. 
So, a slab class of 2 accommodates 6898 chunks each 152 bytes in size.  This slab class stores key-value pairs whose size is between 120 and 152 bytes. The largest slab class uses a chunk size that accommodates the entire slab.

When storing a key-value pair, $k_i$-$v_i$, the server identifies the size required to
store $k_i$-$v_i$ along with some meta-data header information. The memory allocation
attempts the following steps, proceeding in order until it is able to satisfy the allocation request:

\begin{enumerate}

\item Replace an expired key-value of the smallest slab class that can
accommodate the entire $k_i$-$v_i$. 

\item Find a free chunk within allocated slabs of that slab class.

\item Allocate a new slab to the matching slab class and assign one of the chunks. 

\item Evict an existing key-value pair using LRU and replace its contents.

\end{enumerate}

A limitation of the slab allocation system is that, once a slab has been
allocated to a particular slab class, it will forever maintain its class 
assignment. The consequence of this rigid assignment is that it may
prevent future requests from storing key-value pairs in the KVS. 
For example, a certain workload may assign all slabs to the slab class 1 (120 bytes).  Subsequently, the workload may change and require chunks of slab class 5 (304 bytes). Since all slabs were already assigned to slab class 1, all requests to store key-value pairs with a slab class of 5 fail. This phenomenon is termed slab calcification and causes a KVS using slab based allocation to under-utilize its available memory.

Twemcache attempts to resolve this calcification limitation by randomly evicting a slab from another class if it is unable to allocate an item. This approach may evict potentially hotly accessed items, impacting the cache hit rate adversely. Additionally, the random slab eviction does not deal with the case when a disproportionately small number of slabs are assigned to the needed slab class. To illustrate, assume only one slab was assigned to the slab class 5 (0.1\% of total memory) and the workload changes such that key-value pairs corresponding to slab class 5 are referenced much more frequently.   All the requests have to compete for chunks in the single slab whereas the remaining cache space is now under-utilized. Since key-value pairs can still be allocated, the random slab eviction does not activate to free up more slabs.

One may address the calcification limitation by separating how memory should be
allocated for the key-value pairs from the online algorithm that decides which
key-value pairs should occupy the available memory.
For example, with a memcached implementation,
one may use a buddy algorithm~\cite{ier96} to manage space in combination with
CAMP (or LRU).
With those caches that run in the address space of an application (e.g., JBoss,
EhCache, KOSAR), the memory manager of the operating system may manage space for
instances of classes that serve as values for application specified keys. 
With these, one may use CAMP to decide which key-value pairs occupy the available
memory.

\section{Conclusion and Future Research}\label{sec:conc}
We have presented a new efficient implementation of GDS called CAMP. CAMP takes 
advantage of rounded priority values so that the underlying data structure 
selects a key-value pair for eviction very efficiently. Moreover, we have shown 
that on typical access patterns, there is no degradation in performance due to 
loss in precision of the priority values. CAMP outperforms LRU as well as Pooled 
LRU in the overall cost of caching key-value pairs in a sequence of requests in 
which the cost to access different key-value pairs vary dramatically. The 
implementation of CAMP in IQ Twemcache shows that the overhead in implementing 
replacement decisions is as efficient as LRU.

Our short term research plans include examining the performance of CAMP in a 
wider variety of settings. 
It would be particularly interesting to test the 
performance of CAMP on real trace data and in realistic deployments. 
Another important direction to explore is the use of admission control policies in 
conjunction with CAMP that also considers variations in key-value sizes and costs.  
This should enhance the performance of CAMP by not inserting unpopular 
key-value pairs that are evicted before their next request.  

More longer term, we are extending CAMP for use with a hierarchical cache 
(using SSD, hard disk, or both).
This includes an investigation of environments that require CAMP to
manage both key-value pairs that pertain to the result of computations and 
fix sized disk pages used by the computations.
We are also investigating a decentralized CAMP in the context of a cooperative
caching framework such as KOSAR~\cite{mitrallc14}.
A challenge here is how to maintain a last replica of a cached
key-value pair
without allowing those that are never accessed again to occupy the KVS indefinitely.

\bibliographystyle{acm}
\bibliography{cit}  % sigproc.bib is the name of the Bibliography in this case
\end{document}